\documentstyle[preprint,aps,epsfig,tighten,floats]{revtex}

\renewcommand{\Re}{\mathop{\rm Re}}
\renewcommand{\Im}{\mathop{\rm Im}}

\begin{document}

\title{
\null\vskip-0.9in{\small\hfill MPI-PhT/99-35\\\hfill August 1999}\vskip0.3in 
CP-Violating Effects in Neutralino Scattering and Annihilation}

\author{Paolo Gondolo$^1$\thanks{Email: gondolo@mppmu.mpg.de} and Katherine
 Freese$^{1,2}$\thanks{Email: ktfreese@umich.edu}}

\address{$^1$Max Planck Institut f\"ur Physik, F\"ohringer Ring 6, 80805
  M\"unchen, Germany}

\address{$^2$Physics Department, University of Michigan, Ann Arbor, MI 48109,
USA}

\maketitle

\begin{abstract}
  We show that in some regions of supersymmetric parameter space, CP violating
  effects that mix the CP-even and CP-odd Higgs bosons can enhance the
  neutralino annihilation rate, and hence the indirect detection rate of
  neutralino dark matter, by factors of $10^6$. The same CP violating effects
  can reduce the neutralino scattering rate off nucleons, and hence the direct
  detection rate of neutralino dark matter, by factors of $10^{-7}$. We study
  the dependence of these effects on the phase of the trilinear coupling $A$,
  and find cases in the region being probed by dark matter searches which are
  experimentally excluded when CP is conserved but are allowed when CP is
  violated.
\end{abstract}

\section{INTRODUCTION}

The nature of the dark matter in the universe is one of the outstanding
questions in astro/particle physics.  One of the favored candidates is the
lightest supersymmetric (SUSY) particle.  Such a particle is weakly interacting
and massive (with mass in the range 1 GeV -- few TeV), and hence is frequently
characterized as a WIMP (weakly interacting massive particle).  In the minimal
supersymmetric standard model (MSSM), the lightest SUSY particle in most cases
is the lightest neutralino, a linear combination of the supersymmetric partners
of the photon, $Z^0$ boson, and neutral-Higgs bosons,
\begin{equation}
  \tilde{\chi}^0_1 = 
  N_{11} \widetilde{B} + N_{12} \widetilde{W}^3 + N_{13} \widetilde{H}_1^0
  + N_{14} \widetilde{H}_2^0
\end{equation}
where $\widetilde B$ and $\widetilde W^3$ are the supersymmetric partners of
the U(1) gauge field $B$ and of the third component of the SU(2) gauge field
$W^3$ that mix to make the photon and $Z^0$ boson. (We will also use the letter
$\chi$ for $\tilde{\chi}^0_1$.)

Much work has been done studying the possibilities for detecting these
particles.  Possibilities include direct detection \cite{goodman}, whereby the
particle interacts with a nucleon in a low temperature detector, and is
identified by the keV of energy it deposits in the nucleon; and indirect
detection, whereby (1) the particles are captured in the Sun or Earth, sink to
the center of the Sun or Earth, annihilate with one another in the core, and
give rise to particles including neutrinos which can be detected by experiments
on the surface of the Earth \cite{earth-sun}, or (2) the particles annihilate
in the galactic halo and produced anomalous components in the flux of cosmic
rays \cite{halo}.  The interaction processes of the lightest SUSY particle are
clearly of great importance in calculations of predicted rates for both direct
and indirect detection.

In this paper we discuss the effect of CP violation on the neutralino
annihilation and scattering cross sections.  The MSSM introduces several new
phases in the theory which are absent in the standard model.  Supplemented by a
universality condition at the grand unification scale, only two of these are
independent.  In this case, one may choose to work in a basis in which the two
non-trivial CP-violating phases reside in $\mu$ and the universal soft
trilinear coupling $A$ of the Higgs fields to the scalar fermions $\tilde f$.
Previously Falk, Ferstl and Olive \cite{falk} have considered the effect on
neutralino cross sections of a nonzero phase of $\mu$, the mixing mass
parameter involving the two Higgs chiral superfields in the superpotential.
Here, on the other hand, we consider the effect on neutralino cross sections of
the case where the soft trilinear scalar couplings $A_f$ are all complex
numbers, where subscript $f$ refers to the quarks. To be specific, we will take
$A \equiv A_t = A_b$ with arbitrary $\arg(A)$, and we take $\Im(\mu)=0$.  In
part of SUSY parameter space we find enhancement of these cross sections, and
hence an increase in direct and indirect detection rates; while in other parts
of parameter space the cross sections are suppressed.

The phase of $A$ enters into the neutralino cross sections in two places: 1)
into the squark masses, and 2) into the Higgs sector.  For example, one of the
processes that contributes to neutralino annihilation is s-channel exchange of
the three neutral Higgs bosons, $h$, $H$, and $A$, into final state fermions.
(see fig.~1).  The first two of these neutral Higgs bosons, $h$
and $H$, are CP even, while $A$ is CP odd.  The new aspect considered here is
the possibility of mixing between the CP-even Higgs scalars ($h$ and $H$) and
the CP-odd scalar $A$.  This mixing was first studied by Pilaftsis
\cite{pilaftsis}, who found that the size of CP violation can be fairly large,
i.e.\ of order one, for a range of kinematic parameters preferred by SUSY.  He
found that a large $HA$ mixing can naturally occur within two-Higgs doublet
models either at the tree level, if one adds softly CP-violating breaking terms
to the Higgs potential, or at one loop, after integrating out heavy degrees of
freedom that break the CP invariance of the Higgs sector, such as heavy
Majorana neutrinos.  In any case, in this paper, we consider the one-loop
effects of $\Im(A)\neq0$ on scattering and annihilation cross sections
relevant to direct and indirect detection.

In Section II we discuss our general approach.  In Section III, we discuss the
squark sector, and in Section IV, the Higgs sector.  In Section V, we discuss
experimental constraints on the parameters.  We present our results in Section
VI.

\section{General Approach}

The minimal supersymmetric standard model provides a well-defined calculational
framework\cite{haber}, but contains at least 106 yet-unmeasured
parameters~\cite{dim95}.  Most of them control details of the squark and
slepton sectors, and can safely be disregarded in dark matter studies. So
similarly to Bergstr\"om and Gondolo~\cite{ber96}, we restrict the number of
parameters to 6 plus one CP violating phase: the ``CP-odd'' scalar mass $m_A$
(which in our CP violating scenario is just a mass parameter), the Higgs mass
parameter $\mu$, the gaugino mass parameter $M_2$ (we impose gaugino mass
unification), the ratio of Higgs vacuum expectation values $\tan\beta$, a
sfermion mass parameter $\widetilde{M}$ (not to be confused with the sfermion
mass, see eqs.~(\ref{sq-assume}) and~(\ref{msquark}) below), and a complex
sfermion mixing parameter $ A \equiv A_t = A_b$ for the third generation (we
set the $A$'s of the first two generations to zero). The phase of $A$ is the
only CP violating phase we introduce besides the standard model CKM phase.

We use the database of points in parameter space built in
refs.~\cite{ber96,eds97,ber98}, setting their $A_b$ equal to $A_t$.  Hence we
explore a substantial fraction of the supersymmetry parameter space, running
through different possible neutralinos as the lightest SUSY particle.  

We modify the squark and Higgs couplings in the neutralino dark matter code
DarkSUSY \cite{darksusy} to include a non-zero phase of $A$. We also add all
diagrams that contribute to neutralino scattering and annihilation and would
vanish when CP is conserved.

To investigate the effects of the phase of $A$, we perform the following
procedure. For each of the 132,887 sets of parameter values in the database, we
run through 50 values of the phase of $A$, so that we effectively explore
$50\times132,887 \sim 6.6 \times 10^6$ models. We loop over a circle with
$\arg(A)$ varying from 0 to $2\pi$.  At each point, we check bounds on the
electric dipole moment, on the Higgs mass, on other particle masses, on the
$b\to s\gamma$ branching ratio, and on the invisible Z width (table I gives a
listing of the bounds we apply).  If any of these bounds are violated, we move
to the next point on the circle.  If all the bounds are satisfied, we calculate
the spin-independent neutralino--proton scattering cross section $\sigma_{\chi
  p}$.  We record the two values of the phase of $A$ where $\sigma_{\chi p}$ is
highest and lowest, respectively, with the bounds satisfied.  Then, once we
have looped through all the possible values for the phase of $A$, we have found
the two points with the maximum enhancement and suppression of the scattering
cross section.  We then compare with the scattering cross section in the case
of no CP violation.

We do the same for the annihilation cross section times relative velocity
$\sigma v$ at relative velocity $v=0$ (we recall that $\sigma \sim 1/v$ as
$v\to 0$). Thus we obtain the values of the phase of $A$ where $\sigma v$ is
maximum and minimum.

\section{Squark Sector} 

The (complex) scalar top and bottom mass matrices can be expressed in
the $(\tilde{q}_L, \tilde{q}_R)$ basis as
\begin{equation}
  {\cal M}^2_{\tilde q} = \pmatrix{
    M_{\widetilde{Q}}^2 + m_q^2 + 
    \left( T_{3q} - e_q \sin^2\theta_W\right)\cos 2\beta m_Z^2
    & m_q\left(A^*_q - \mu R_q\right)
    \cr 
    m_q\left(A_q - \mu^* R_q\right)  &
    M_{\widetilde{R}}^2 + m_q^2 + 
    e_q \sin^2\theta_W m_Z^2 \cos 2\beta \cr} , 
\end{equation}
where $q=t$ or $b$; $e_t={2 \over 3}$; $e_b = -{1 \over 3}$; $T_{3t} = {1 \over
  2}$; $T_{3b} = -{1 \over 2}$; $R_t=\cot\beta$, $R_b=\tan\beta$; and
$M_{\widetilde{R}}^2=M_{\widetilde{U}}^2$ $[M_{\widetilde{D}}^2]$ for $t$
$[b]$. We set 
\begin{equation}
\label{sq-assume}
M_{\widetilde{Q}} = M_{\widetilde{U}} = M_{\widetilde{D}} = \widetilde{M},
\end{equation}
our sfermion mass parameter. Even in the case of no CP
violation, when both $\mu$ and $A$ are real, there is mixing between the
squarks, and this matrix must be diagonalized to find the mass eigenstates.
Here we take $A$ to be complex.  Then we obtain the mass eigenstates $\tilde
q_1, \tilde q_2$ from the weak eigenstates $\tilde q_L, \tilde q_R$ through the
rotation
\begin{equation}
\label{sqmixing}
\pmatrix{\tilde q_1\cr \tilde q_2\cr} = \pmatrix{\cos\theta_{\tilde{q}} &
\sin\theta_{\tilde{q}} e^{i \gamma_{\tilde{q}}}\cr 
- \sin\theta_{\tilde{q}}& 
\cos\theta_{\tilde{q}} e^{i\gamma_{\tilde{q}}}\cr} \pmatrix{\tilde
q_L\cr \tilde q_R\cr} , 
\end{equation}
where $\gamma_{\tilde{q}} = \arg(A^*_q-\mu R_q)$ and the rotation angle
$\theta_{\tilde{q}}$ ($-\pi/4 \le \theta_{\tilde{q}} \le \pi/4$) may be
obtained by
\begin{equation}
\tan(2\theta_{\tilde{q}}) = 
{2 m_q | A^*_q - \mu R_q|\over M^2_{\widetilde{R}} - M^2_{\widetilde{Q}}
+  (2 e_q \sin^2\theta_W - T_{3q})  m_Z^2 \cos 2\beta }.
\end{equation}
The masses of $\tilde q_1$ and $\tilde q_2$ are then given by
\begin{eqnarray}
m^2_{\tilde q_{1,2}} & = &{1\over 2} \Biggl\{ 
M^2_{\widetilde{Q}} + M^2_{\widetilde{R}} + T_{3q} m_Z^2 \cos 2\beta
\nonumber\\
&& \pm \mathop{\rm sign}(\theta_{\tilde{q}})
\sqrt{\Bigl[ M^2_{\widetilde{R}} - M^2_{\widetilde{Q}}
+  (2 e_q \sin^2\theta_W - T_{3q})  m_Z^2 \cos 2\beta\Bigr]^2 + 
4m_q^2 \Bigl| A^*_q - \mu R_q\Bigr|^2}\Biggr\}.
\label{msquark}
\end{eqnarray}
The $+$ sign is for $\tilde q_1$ and the $-$ sign for $\tilde q_2$. 

The mixing in eq.~(\ref{sqmixing}) also modifies the squark couplings to the
neutralino and the corresponding quark. Writing the relevant interaction term
as
\begin{equation}
  {\cal L}_{\rm int} = \tilde{q}_i \overline{\chi}
 \left( g^L_{\tilde{q}_i\chi q} P_L + g^R_{\tilde{q}_i\chi q} P_R
 \right) q + \hbox{h.c.} ,
\end{equation}
with $P_L = (1-\gamma_5)/2$, $P_R = (1+\gamma_5)/2$, and $i=1,2$, we have 
\begin{equation}
\label{sq-x-q}
\pmatrix{ g^K_{\tilde{q}_1\chi q} \cr  g^K_{\tilde{q}_2\chi q} } 
=
\pmatrix{\cos\theta_{\tilde{q}} &
\sin\theta_{\tilde{q}} e^{i \gamma_{\tilde{q}}}\cr 
- \sin\theta_{\tilde{q}}& 
\cos\theta_{\tilde{q}} e^{i\gamma_{\tilde{q}}}\cr}
 \pmatrix{g_{KL}\cr g_{KR}\cr} 
\end{equation}
where $K=L,R$,
\begin{equation}
g_{LL} = -\sqrt{2} \left(T_{3q}gN_{12}+(e_q-T_{3q})g' N_{11}\right),
\qquad\qquad
g_{RR} = \sqrt{2} e_q g' N_{11},
\end{equation}
and
\begin{equation}
g_{LR} = g_{RL} = -{g m_u N_{14}\over \sqrt{2} m_W \sin\beta}
\end{equation}
for the up-type quarks, 
\begin{equation}
g_{LR} = g_{RL} = -{g m_d N_{13}\over \sqrt{2} m_W \cos\beta}
\label{sq-x-q-last}
\end{equation}
for the down-type quarks.

The expressions in this section apply to sleptons provided up-type (s)quarks is
replaced with (s)neutrinos and down-type (s)quarks with charged (s)leptons.

\section{Higgs Sector}

\subsection{Higgs masses}
We evaluate the Higgs boson masses in the effective potential approach. The
radiatively corrected Higgs boson mass matrix can be written as
\begin{equation}
\label{higgsmass}
  {\cal M}^2 = \left( \matrix{ 
      {m_Z^2 \cos^2\beta + m_A^2 \sin^2\beta + \Delta_{11} } &
      {-(m_A^2+m_Z^2)\sin\beta\cos\beta + \Delta_{12} } & 
      \Delta_{13}
      \cr
      {-(m_A^2+m_Z^2)\sin\beta\cos\beta + \Delta_{21} } &
      {m_Z^2 \sin^2\beta + m_A^2 \cos^2\beta + \Delta_{22} } & 
      \Delta_{23}
      \cr
      \Delta_{31} & \Delta_{32} & m_A^2 
      } \right) 
\end{equation}
in the basis $H_1$, $H_2$, $H_3$. Here $\Delta_{ij}=\Delta_{ji}$ are the
radiative corrections coming from quark and squark loops, with $\Delta_{13}$
and $\Delta_{23}$ arising from CP violation. We take $\Delta_{11}$,
$\Delta_{12}$, and $\Delta_{22}$ from ref.~\cite{ellis+}.
\begin{eqnarray}
  \Delta_{11} &=& {3g^2\over 16\pi^2m_W^2} \left[ {m_b^4\over\cos^2\beta}
\left( \ln{m^2_{\tilde{b}_1}m^2_{\tilde{b}_2}\over m_b^4} +
2 Z_b \ln {m^2_{\tilde{b}_1}\over m^2_{\tilde{b}_2}} \right) \right.
\nonumber\\ && \left.\qquad \qquad\quad +
{m_b^4\over \cos^2\beta} Z_b^2 
g(m^2_{\tilde{b}_1},m^2_{\tilde{b}_2}) +
{m_t^4\over \sin^2\beta} W_t^2 
g(m^2_{\tilde{t}_1},m^2_{\tilde{t}_2})  \right] ,
  \\
  \Delta_{22} &=& {3g^2\over 16\pi^2m_W^2} \left[ {m_t^4\over\sin^2\beta}
\left( \ln{m^2_{\tilde{t}_1}m^2_{\tilde{t}_2}\over m_t^4} +
2 Z_t \ln {m^2_{\tilde{t}_1}\over m^2_{\tilde{t}_2}} \right) \right.
\nonumber\\ && \left.\qquad \qquad\quad +
{m_t^4\over \sin^2\beta} Z_t^2 
g(m^2_{\tilde{t}_1},m^2_{\tilde{t}_2}) +
{m_b^4\over \sin^2\beta} W_b^2 
g(m^2_{\tilde{b}_1},m^2_{\tilde{b}_2})  \right] ,
\\
  \Delta_{12} &=& {3g^2\over 16\pi^2m_W^2} \left[ 
    {m_t^4\over \sin^2\beta} \, W_t \,
    \left( \ln {m^2_{\tilde{t}_1}\over m^2_{\tilde{t}_2}} + Z_t 
      g(m^2_{\tilde{t}_1},m^2_{\tilde{t}_2}) \right)
      \right. \nonumber \\ && \qquad\qquad\quad \left. +
    {m_b^4\over \cos^2\beta} \, W_b \,
    \left( \ln {m^2_{\tilde{b}_1}\over m^2_{\tilde{b}_2}} + Z_b 
      g(m^2_{\tilde{b}_1},m^2_{\tilde{b}_2}) \right) \right] ,
\end{eqnarray}
where
\begin{eqnarray}
  W_q &=& { \Re(\mu A_q) - |\mu|^2 R_q \over 
    m^2_{\tilde{q}_2} - m^2_{\tilde{q}_1} }  ,
  \\
  Z_q &=& { |A_q|^2 - \Re(\mu A_q) R_q \over 
    m^2_{\tilde{q}_2} - m^2_{\tilde{q}_1} } ,
  \\
  g(m_{1}^2, m_{2}^2)
   &=& 2 - { m^2_{1} + m^2_{2} \over
    m^2_{1} -  m^2_{2} } 
  \ln {m^2_{1}\over m^2_{2}} .
\end{eqnarray}
We have rewritten $\Delta_{13}$ and $\Delta_{23}$ from ref.~\cite{pilaftsis} in
a way that shows their proportionality to $\Im(\mu A)$.
\begin{eqnarray}
\Delta_{k3} & = & {3\over 16\pi^2}\sum_q g_{A\tilde{q}_1 \tilde{q}_1}
\Biggl\{ {1\over 2} \left(g_{H_k \tilde q_L \tilde q_L} + g_{H_k
\tilde q_R \tilde q_R}\right) \log {m^2_{\tilde q_1}\over m^2_{\tilde q_2}}\\
&+ & \left[ 
\sin 2 \theta_q \Re(e^{i\gamma_q} g_{H_k\tilde q_R \tilde q_L}) +
{1\over 2} \cos 2 \theta_q \left(g_{H_k \tilde q_L \tilde q_L} - g_{H_k
\tilde q_R \tilde q_R}\right) \right]
g\!\left(m_{\tilde q_1}^2, m_{\tilde q_1}^2\right)\Biggr\} ,
\end{eqnarray}
where the couplings of the Higgs bosons to the squarks are
\begin{eqnarray}
g_{A \tilde t_1 \tilde t_1} & = & - {g m_t^2\over m_W \sin^2\beta}\
{\Im(\mu A_t)\over m_{\tilde t_1}^2 - m_{\tilde t_2}^2} ,
\\
g_{A \tilde b_1 \tilde b_1} & = & - {g m_b^2\over m_W \cos^2\beta}\
{\Im(\mu A_b)\over m_{\tilde b_1}^2 - m_{\tilde b_2}^2} ,
\\
g_{H_1 \tilde t_L \tilde t_L} & = & -
{g m_Z\over \cos\theta_W} \left(T_{3t} - e_t \sin^2\theta_W\right)
\cos\beta ,
\\
g_{H_1 \tilde t_R \tilde t_R} & = & - 
{g m_Z\over \cos\theta_W} e_t  \sin^2\theta_W
\cos\beta ,
\\
g_{H_1 \tilde t_R \tilde t_L} & = & {g m_t \mu^*\over 2 m_W \sin\beta} ,
\\
g_{H_1 \tilde b_L \tilde b_L} & = & -
{g m_b^2\over m_W \cos\beta} - 
{g m_Z\over \cos\theta_W} \left(T_{3b} - e_b \sin^2\theta_W\right)\cos\beta ,
\\
g_{H_1 \tilde b_R \tilde b_R} & = & -
{g m_b^2\over m_W \cos\beta} -
{g m_Z\over \cos\theta_W} e_b \sin^2\theta_W\cos\beta ,
\\
g_{H_1 \tilde b_R \tilde b_L} & = & -{g m_b A_b\over 2 m_W \cos\beta} ,
\\
g_{H_2 \tilde t_L \tilde t_L} & = & - {g m_{b^2}\over m_W \cos\beta} +
{g m_Z\over \cos\theta_W} \left(T_{3t} - e_t \sin^2 \theta_W\right)
\sin\beta ,
\\
g_{H_2 \tilde t_R \tilde t_R} & = & 
- {g m_b^2\over m_W \cos\beta} + {g m_Z\over \cos\theta_W} e_t \sin^2
\theta_W \sin \beta ,
\\
g_{H_2 \tilde t_R \tilde t_L} & = & - {g m_t A_t\over 2 m_W \sin\beta} ,
\\
g_{H_2 \tilde b_L \tilde b_L} & = & 
 {g m_Z\over \cos\theta_W} \left(T_{3b} - e_b \sin^2 \theta_W\right)
\sin\beta ,
\\
g_{H_2 \tilde b_R \tilde b_R} & = & 
{g m_z\over \cos\theta_W}  e_b \sin^2 \theta_W \sin\beta ,
\\
g_{H_2 \tilde b_R \tilde b_L} & = & {g m_b \mu^*\over 2 m_W \cos\beta} .
\end{eqnarray}

Neglecting D terms, as we should for consistency with the CP even part and the
vertices in our effective potential approach, the corrections $\Delta_{13}$ and
$\Delta_{23}$ simplify to
\begin{eqnarray}
\Delta_{13} & = & {3 g^2 \over 16 \pi^2 m_W^2} \left[ 
{m_b^4\over \cos^3\beta} \, X_b  \, 
\left( \ln {m^2_{\tilde{b}_1}\over m^2_{\tilde{b}_2}} + 
Z_b g(m^2_{\tilde{b}_1},m^2_{\tilde{b}_2}) \right) 
\right. \nonumber \\ && \qquad\qquad\quad \left. + 
{m_t^4\over \sin^3\beta} \, X_t W_t \, 
g(m^2_{\tilde{t}_1},m^2_{\tilde{t}_2}) \right] ,
\\
\Delta_{23} & = & {3 g^2 \over 16 \pi^2 m_W^2} \left[ 
{m_t^4\over \sin^3\beta} \, X_t  \, 
\left( \ln {m^2_{\tilde{t}_1}\over m^2_{\tilde{t}_2}} + 
Z_t g(m^2_{\tilde{t}_1},m^2_{\tilde{t}_2}) \right) 
\right. \nonumber \\ && \qquad\qquad\quad \left. + 
{m_b^4\over \cos^3\beta} \, X_b W_b \, 
g(m^2_{\tilde{b}_1},m^2_{\tilde{b}_2}) \right] ,
\end{eqnarray}
with
\begin{equation}
X_q = { \Im(\mu A_q) \over m^2_{\tilde q_1} - m^2_{\tilde q_2} } .
\end{equation}

The key thing to notice is that the $\Delta_{k3}$ self-energies
are proportional to Im($\mu A)$.  For $\mu$ real, they are hence proportional
to $\Im(A)$.  

We use the effective potential approach to obtain the Higgs masses and
couplings. The Higgs mass eigenstates $h_i$ ($i=1,2,3$) are obtained by
diagonalizing the Higgs mass matrix including radiative corrections in
eq.~(\ref{higgsmass}) through the orthogonal Higgs mixing matrix $O$ as
\begin{equation}
H_i = O_{ij} h_j
\end{equation}
In practice, it is convenient to implement the diagonalization in two steps, to
separate the CP violating contributions. First we diagonalize the ``CP-even''
part through
\begin{equation}
H_i = O^0_{ij} \Phi_j ,
\end{equation}
where $\Phi_i = H, h, A$ for $i=1,2,3$ respectively.
The matrix $O^0$ would be the Higgs mixing matrix in absence of CP violation
\begin{equation}
O^0 = \pmatrix{\cos \alpha & - \sin\alpha & 0 \cr
\sin\alpha & \cos \alpha & 0 \cr 0 & 0 & 1 \cr} ,
\end{equation}
with
\begin{equation}
\tan(2\alpha) = { 2 {\cal M}^2_{12} \over {\cal M}^2_{11} - {\cal M}^2_{22} } .
\end{equation}
Then we further rotate to the mass eigenstates with an orthogonal matrix $O'$
as
\begin{equation}
\Phi_i = O'_{ij} h_j
\end{equation}
with $O' = O O^{0T}$. This two step procedure allows for a rapid introduction
of CP violating mixing angles for the Higgs sector in the DarkSUSY code.

\subsection{Higgs couplings}

We will include CP violating effects by rotating couplings of Higgs particles
to other particles as described in this section. In the effective potential
approach we neglect vertex corrections. This incorporates the dominant
corrections of ${\cal O}(g^2m_t^4/m_W^4)$, and neglects corrections of ${\cal
  O}(g^2m_t^2/m_W^2)$.

There are terms in the Lagrangian that couple the Higgs particles to other
particles that are linear in the Higgs fields, for example
\begin{equation}
g_{\Phi_i qq} \Phi_i \bar q q = g_{h_i qq} O'_{ji} h_j \bar q q.
\end{equation}
Terms of this type include coupling to fermions, as shown above, and also terms
such as $g_{WH^+\Phi_i} W H^+ \Phi_i$.  We will define rotated couplings via
\begin{equation}
g_{h_i ab} = O'_{ij} g_{\Phi_j ab} 
\end{equation}
where $a$ and $b$ stand for the appropriate particle name.

Those terms with two Higgs bosons in them, such as 
\begin{equation}
g_{Z \Phi_3 \Phi_i} Z
\Phi_3 \partial_\mu \Phi_i = g_{Z \Phi_3 \Phi_i} O'_{k3} O'_{ji} Z h_k
\partial_\mu h_j,
\end{equation}
must have the couplings rotated with two multiplications
by $O'$, e.g., 
\begin{equation}
g_{Z h_k h_j} = g_{Z A \Phi_i} O'_{ji} O'_{k3} -
(k \leftrightarrow j).
\end{equation}
Note that, in this particular term, the appropriate
antisymmetry properties are maintained, and $i$ takes on values 1 or 2 only.

We have carefully rotated all couplings involving one, two, or three Higgs
bosons.  It is these rotated couplings that we use in the numerical code.
(i.e.\ we replace the ordinary Higgs couplings with these rotated couplings.)

As an example, we give the Higgs--quark and Higgs--neutralino vertices that
appear in the neutralino--proton spin-independent cross section.
\begin{eqnarray}
\label{h-q-q}
g_{h_iuu} &=& - {g m_u \over 2 m_W \sin\beta} 
\left( O'_{i1} \sin\alpha + O'_{i2} \cos\alpha + O'_{i3} i \cos\beta \right),
\\
g_{h_idd} &=& - {g m_d \over 2 m_W \cos\beta} 
\left( O'_{i1} \cos\alpha - O'_{i2} \sin\alpha + O'_{i3} i \sin\beta \right) ,
\\
g_{h_i\chi_m\chi_n} &=& {1\over2} ( g N^*_{m2} - g' N^*_{m1} ) 
\left[ N_{n3}^* 
  \left(-O'_{i1}\cos\alpha+O'_{i2}\sin\alpha+O'_{i3}i\sin\beta \right) + 
\right. \nonumber\\ && \left. \qquad\qquad\qquad\qquad
       N_{n4}^*
  \left(O'_{i1}\sin\alpha+O'_{i2}\cos\alpha-O'_{i3}i\cos\beta \right)
\right] + (m\leftrightarrow n) .
\label{h-x-x}
\end{eqnarray}
Here $u$ stands for down-type quarks and neutrinos, $d$ stands for up-type
quarks and charged leptons.

\section{Experimental Bounds}

\subsection{Bounds on masses}

We impose experimental bounds on the invisible width of the $Z^0$ boson,
$\Gamma_Z^{\rm inv}$, and on particle masses as listed in table I.

Since the $h$, $H$, and $A$ are rotated into new mass eigenstates bosons, we
use the most model independent constraint on the neutral Higgs masses: we take
$m_{h_i} > 82.5$ GeV.  This constraint was reported by the ALEPH group
\cite{gao99} at the 95\% C.L.  as a bound on all Higgs masses, independent of
sin$^2(\beta - \alpha)$.  Note that this bound, which is a 10\% improvement
over previous bounds, renders the cross section for direct detection of SUSY
particles smaller by a factor of two.  This suppression arises because the
dominant contribution to the scattering cross section is via Higgs exchange and
scales as $\sigma_{\chi p} \propto 1/m_{h_i}^4$.

\subsection{Bounds on CP violation}

We impose bounds on the branching ratio ${\rm BR}(b\to s\gamma)$, and on the
electric dipole moments of the electron and of the neutron $d_e$ and $d_n$.

For  ${\rm BR}(b\to s\gamma)$ we use the expressions in ref.~\cite{bertolini},
with inclusion of the one-loop QCD corrections.

Since we assume that the only new CP violating phase is that of $A$, the
leading contribution to the electric dipole moment (EDM) arises at two-loops
\cite{chang}.  Chang, Keung, and Pilaftsis \cite{chang} have calculated 
two-loop contributions to the electric dipole moment (EDM) which originate from
the potential CP violation due to a nonzero phase of $A$. We rewrite them
showing explicitly their dependence on $\Im(\mu A)$. They find the electric and
chromo-electric EDM of a light fermion $f$ at the electroweak scale as
\begin{equation}
(d^E_f)_{EW} = e e_f {3 \alpha_{\rm em} \over 64 
\pi^3} {R_f m_f \over m_A^2} \sum_{q=t,b} \xi_q e_q^2
\left[F\!\left({m_{\tilde{q}_1}^2 \over m_A^2}\right)
- F\!\left({m_{\tilde{q}_2}^2 \over m_A^2}\right) \right] ,
\end{equation}
\begin{equation}
(d^C_f)_{EW} = g_s e_f {\alpha_{s} \over 128
\pi^3} {R_f m_f \over m_A^2} \sum_{q=t,b} \xi_q 
\left[F\!\left({m_{\tilde{q}_1}^2 \over m_A^2}\right)
- F\!\left({m_{\tilde{q}_2}^2 \over m_A^2}\right) \right] ,
\end{equation}
where $\alpha_{\rm em} = e^2/(4\pi)$ is the electromagnetic fine structure
constant, $\alpha_s = g_s^2/(4\pi)$ is the strong coupling constant, all the
kinematic parameters must be evaluated at the electroweak scale $m_Z$, $e_i$ is
the electric charge of particle $i$, $R_f=\tan\beta$ for $f=u,c,t$,
$R_f=\cot\beta$ for $f=e,\mu,\tau,d,s,b$, and $F(z)$ is a two-loop function
given by
\begin{equation}
F(z) = \int_0^1 dx {x(1-x) \over z-x(1-x)} \ln
\left[{x(1-x) \over z} \right] .
\end{equation}
The EDM of the neutron can then be estimated by a naive dimensional analysis
\cite{manohar,ibrahimnath} as
\begin{equation}
d_n = \eta^E {1\over 3} \left( 4 d^E_d - d^E_u \right) + 
\eta^C {e\over 12\pi } \left( 4 d^C_d - d^C_u \right) .
\end{equation}
We take the numerical values $\eta^E=1.53$ and $\eta^C=3.4$ \cite{ibrahimnath}.

The CP violating quantities $\xi_t$ and $\xi_b$ are given by
\begin{equation}
\xi_t = -{g m_t^3 \Im(\mu A_t) \over 2 m_W^2 \sin^2\!\beta (m_{{\tilde t}_1}^2
-m_{{\tilde t}_2}^2)} 
\end{equation}
and
\begin{equation}
\xi_b = -{g m_b^3 \Im(\mu A_b) 
\over 2 m_W^2 \cos^2\!\beta (m_{{\tilde b}_1}^2 - m_{{\tilde b}_2}^2) } .
\end{equation}
As an upper bound to the contribution to the measured value of the electron EDM
we take $|d_e| < 0.4 \times 10^{-26} e$cm \cite{commins94}.  The bound on the
neutron EDM is $|d_n| < 1.79 \times 10^{-25} e$cm \cite{altarev96}.  We keep
only models that satisfy these bounds.

\begin{table*}[ht]
\label{tab:1}
\begin{center}
\begin{tabular}{|l l|}
\hline \hfil Bound \hfill & \hfil Ref. \hfill \\ \hline $
\Gamma_Z^{\rm inv} < 502.4 {\rm MeV}
$ & \cite{pdg99} \\ \hline $
m_{H^\pm} > 59.5{\rm GeV} 
$ & \cite{abbiendi99} \\ $
m_{h_i} > 82.5 {\rm GeV}
$ & \cite{gao99} \\ \hline $
m_{\tilde{\chi}^+_1} > 91 {\rm GeV}
  \;\hbox{if}\; m_{\tilde{\chi}^0_1}-m_{\tilde{\chi}^+_2} > 4 {\rm GeV} 
$ & \cite{carr98} \\ $
m_{\tilde{\chi}^+_1} > 64 {\rm GeV} 
    $ \hbox{if} $  m_{\tilde{\chi}^0_1} > 43 {\rm GeV}
    $ \hbox{and} $  m_{\tilde{\chi}^+_2} > m_{\tilde{\chi}^0_2}
$ & \cite{acciarri96} \\ $ 
m_{\tilde{\chi}^+_1} > 47 {\rm GeV} 
    $ \hbox{if} $  m_{\tilde{\chi}^0_1} > 41 {\rm GeV} 
$ & \cite{decamp92} \\ $ 
m_{\tilde{\chi}^+_2} > 99 {\rm GeV}
$ & \cite{hidaka91} \\ \hline $
m_{\tilde{\chi}^0_1} > 23{\rm GeV}  $ \hbox{if} $  \tan\beta>3
$ & \cite{acciarri95} \\ $
m_{\tilde{\chi}^0_1} > 20{\rm GeV}  $ \hbox{if} $  \tan\beta>2
$ & \cite{acciarri95} \\ $
m_{\tilde{\chi}^0_1} > 12.8{\rm GeV}  $ \hbox{if} $  m_{\tilde{\nu}}<200
  {\rm GeV} 
$ & \cite{buskulic96} \\ $
m_{\tilde{\chi}^0_1} > 10.9{\rm GeV} 
$ & \cite{acciarri98} \\ $
m_{\tilde{\chi}^0_2} > 44{\rm GeV} 
$ & \cite{abbiendi99b} \\ $
m_{\tilde{\chi}^0_3} > 102{\rm GeV} 
$ & \cite{abbiendi99b} \\ $
m_{\tilde{\chi}^0_4} > 127{\rm GeV} 
$ & \cite{acciarri95} \\ \hline $ 
m_{\tilde{g}} > 212{\rm GeV}  $ \hbox{if} $  m_{\tilde{q}_k} <
               m_{\tilde{g}}
$ & \cite{abachi95} \\ $
m_{\tilde{g}} > 162{\rm GeV}
$ & \cite{abe97} \\ \hline $
m_{\tilde{q}_k} > 90 {\rm GeV}  $ \hbox{if} $  m_{\tilde{g}} <410{\rm GeV}
$ & \cite{abe92} \\ $ 
m_{\tilde{q}_k} > 176 {\rm GeV}  $ \hbox{if} $  m_{\tilde{g}} <300{\rm GeV}
$ & \cite{abachi95} \\ $
m_{\tilde{q}_k} > 224 {\rm GeV}  $ \hbox{if} $  m_{\tilde{g}}>m_{\tilde{g}}
$ & \cite{abe96} \\ $
m_{\tilde{e}} > 78{\rm GeV}  $ \hbox{if} $  m_{\tilde{\chi}^0_1}<73{\rm
  GeV} 
$ & \cite{barate98} \\ $
m_{\tilde{\mu}} > 71{\rm GeV}  $ \hbox{if} $  m_{\tilde{\chi}^0_1}<66{\rm
  GeV} 
$ & \cite{barate98} \\ $
m_{\tilde{\tau}} > 65{\rm GeV}  $ \hbox{if} $  m_{\tilde{\chi}^0_1}<55{\rm
  GeV} 
$ & \cite{barate98} \\ $
m_{\tilde{\nu}} > 44.4{\rm GeV}
$ & \cite{pdg99} \\ \hline $
1 \times 10^{-4} < {\rm BR}(b\to s\gamma) < 4 \times 10^{-4}
$ & \cite{pdg99} \\ $
|d_e| < 0.4 \times 10^{-26} e{\rm cm}
$ & \cite{commins94} \\ $
|d_n| < 1.79 \times 10^{-25} e{\rm cm}
$ & \cite{altarev96} \\ \hline
\end{tabular}
\end{center}
\caption{Experimental bounds we use in this paper. We do not include
  cosmological bounds nor bounds from dark matter searches.}
\end{table*}

\section{Scattering cross section}

The neutralino--proton scattering cross section for spin-independent
interactions can be written as
\begin{equation}
\sigma_{\chi p} = { G^2_{\chi p} \mu^2_{\chi p} \over \pi },
\end{equation}
where $\mu_{\chi p} = m_\chi m_p/(m_\chi+m_p) $ is the reduced
neutralino--proton mass, and
\begin{equation}
G_{\chi p} =  \sum_q {f_q m_p \over m_q} \left[ 
  \sum_{i=1}^{3} {\Re(g_{h_i\chi\chi}) \Re(g_{h_i q q}) \over m_{h_i}^2 } 
  - {1\over 2} \sum_{k=1}^2 
  { \Re( g^L_{\tilde{q}_k \chi q} g^{R*}_{\tilde{q}_k \chi q} )
    \over m_{\tilde{q}_k}^2 
    } \right] .
\label{Gxp}
\end{equation}
The sum over $q$ runs over all quarks. The coupling constants are given in
eqs.~(\ref{sq-x-q},\ref{h-q-q}--\ref{h-x-x}). We take \cite{gas91+}
\begin{equation}
f_u = 0.023, \quad f_d = 0.034, \quad f_s = 0.14, \quad
f_c = f_b = f_t = 0.0595,
\end{equation}
and
\begin{equation}
\begin{array}{l}
m_u = 5.6{\rm MeV}, \quad m_d = 9.9{\rm MeV}, \quad m_s = 199{\rm MeV}, \\
m_c = 1.35{\rm GeV}, \quad m_b = 5{\rm GeV}, \quad m_t = 175{\rm GeV}.
\end{array}
\end{equation}

Notice that only the real part of the couplings of the Higgs and neutralinos to
Higgs bosons in eq.~(\ref{h-q-q}--\ref{h-x-x}) enter the scattering cross
section. Since both $g_{Aqq}$ and $g_{A\chi\chi}$ are purely imaginary (because
$\Im(\mu)=0$), introducing a phase in $A$ cannot possibly enhance the Higgs
couplings in eq.~(\ref{Gxp}). Similarly, the neutralino--squark--quark
couplings can only be suppressed for $\Im(A) \ne 0$.
However, enhancements to the scattering cross section can still come from the
Higgs or squark masses in the denominator in eq.~(\ref{Gxp}).

\section{Annihilation cross section}

The neutralino--neutralino annihilation cross section times relative velocity
$\sigma v$ is relevant for neutralino annihilations in the center of the Earth
and Sun and in the galactic halo. An enhancement in $\sigma v$ may lead to a
higher annihilation signal from the Earth when the capture of neutralinos in
the core has not yet reached equilibrium with their self--annihilation. An
increased $\sigma v$ gives directly an increased intensity of positron,
antiproton, and gamma-ray fluxes from neutralino annihilation in the galactic
halo.

The neutralino annihilation cross section also determines the relic density of
neutralinos. In this case, there are important contributions at $v\ne 0$
(p-waves, etc.) in large regions of the supersymmetric parameter space. Due to
the excessive computational cost of obtaining the relic density in presence of
CP violation, in this paper we consider only the $v=0$ case, and postpone the
study of the effect of CP violating phases on the neutralino relic density. The
enhancements and suppressions of $\sigma v$ at $v=0$ that we obtain in the
following are indications of analogous enhancements and suppressions in the
neutralino relic density.

The annihilation cross section at $v=0$ includes the following contributions
\begin{equation}
\sigma v = \left[ \sum_f \sigma_{\overline{f}f} + \sigma_{W^+W^-} + 
\sigma_{ZZ} + \sigma_{H^+W^-} + \sigma_{H^-W^+} + \sum_{i=1}^3 \sigma_{h_iZ} + 
 \sum_{ij=1}^3 \sigma_{h_ih_j} 
\right] v
\end{equation}
where $\sigma_{XY}$ refers to the annihilation channel $\chi\chi \to XY$, which
is open when $2 m_\chi \ge m_X+m_Y$.

The annihilation cross section in each channel can be written in terms of
helicity amplitudes ${\cal A}$ as
\begin{equation}
\label{helic}
\sigma_{XY} v = {\lambda_{XY} \over 128 \pi m_\chi^2} 
\sum_{\rm helicities} \Bigl| {\cal
    A} \Bigr|^2 
\end{equation}
where the amplitudes are normalized as in ref.~\cite{pdg99} and
\begin{equation}
\lambda_{XY} = \sqrt{ \left[ 1 - {(m_X+m_Y)^2\over 4 m_\chi^2} \right] 
  \left[ 1 - {(m_X-m_Y)^2\over 4 m_\chi^2} \right] }.
\end{equation}
The DarkSUSY code already includes analytic expressions for each helicity
amplitude required in eq.~(\ref{helic}), with arbitrary complex couplings
between the particles. Hence once we have rotated all vertices as described in
sect.~IV, and have added all annihilation diagrams that vanish when CP is
conserved (e.g.\ the s-channel exchange of all Higgs bosons), the annihilation
cross section including CP violation is automatically calculated correctly by
DarkSUSY.

For future reference, we list the individual contributions to the annihilation
cross section including terms that violate CP.
\begin{eqnarray}
\label{xxff}
  \sigma_{\overline{f}f} v &=& {N_f \lambda_{ff} m_{\chi}^2 \over 32 \pi } 
  \left| 
    \sum_{i=1}^3 { 4 \Im(g_{h_iff}) \Im(g_{h_i\chi_1\chi_1})
      \over m_{h_i}^2-4m_\chi^2  -im_{h_i}\Gamma_{h_i}} 
    + { 4 g^A_{Zff} \Re(g_{Z\chi_1\chi_1}) (m_f /m_\chi)
      \over m_Z^2} +
    \nonumber \right. \\ && \qquad\qquad\;\; \left. +
    \sum_{s=1}^2 { (|g^R_{\tilde{f}_s\chi f}|^2 +
      |g^L_{\tilde{f}_s\chi f}|^2) ( m_f/m_{\chi} ) + 
      2 \Re(g^L_{\tilde{f}_s\chi f} g^{R*}_{\tilde{f}_s\chi f}) \over 
      m_{\tilde{f}_s}^2 + m_\chi^2 - m_f^2 
      -im_{\tilde{f}_s}\Gamma_{\tilde{f}_s}} 
  \right|^2 + 
  \nonumber \\ &+& {N_f \lambda_{ff}^3 m_{\chi}^2 \over 32 \pi } 
  \left| 
    \sum_{i=1}^3 { 4 i \Re(g_{h_iff}) \Im(g_{h_i\chi_1\chi_1})
      \over m_{h_i}^2-4m_\chi^2  -im_{h_i}\Gamma_{h_i}} +
    \sum_{s=1}^2 { |g^R_{\tilde{f}_s\chi f}|^2 -
      |g^L_{\tilde{f}_s\chi f}|^2 \over 
      m_{\tilde{f}_s}^2 + m_\chi^2 - m_f^2 
      -im_{\tilde{f}_s}\Gamma_{\tilde{f}_s}}
  \right|^2 ,
\\
  \sigma_{W^+W^-} v &=& {\lambda_{WW} \over 8 \pi} \left|
    \sum_{c=1}^2 { \lambda_{WW} m_\chi 
      ( |g^R_{W\chi \tilde{\chi}^+_c}|^2+|g^L_{W\chi \tilde{\chi}^+_c}|^2 ) +
      2 i m_{\tilde{\chi}^+_c} 
      \Im(g^L_{W\chi \tilde{\chi}^+_c}g^{R*}_{W\chi \tilde{\chi}^+_c})
      \over m^2_{\tilde{\chi}^+_c} + m_\chi^2 - m_W^2 
      -im_{\tilde{\chi}^+_c}\Gamma_{\tilde{\chi}^+_c}} 
  \right|^2 +
  \nonumber \\ &+&
  {\lambda_{WW} \over 4 \pi} \left|
    \sum_{c=1}^2 { [2(m_\chi/m_W)^2-1] m_{\tilde{\chi}^+_c}
      \Im(g^L_{W\chi \tilde{\chi}^+_c}g^{R*}_{W\chi \tilde{\chi}^+_c}) 
      \over m^2_{\tilde{\chi}^+_c} + m_\chi^2 - m_W^2 
      -im_{\tilde{\chi}^+_c}\Gamma_{\tilde{\chi}^+_c}} 
  \right|^2 ,
\label{sww}
\\
\label{szz}
  \sigma_{ZZ} v &=& {\lambda_{WW} \over 16 \pi} \left|
    \sum_{n=1}^4 { 2 \lambda_{ZZ} m_\chi |g_{Z\chi\chi_n}|^2 -
      2 i m_{\chi_n} \Im(g^2_{Z\chi\chi_n})
      \over m^2_{\chi_n} + m_\chi^2 - m_Z^2 
      -im_{\chi_n}\Gamma_{\chi_n}} 
  \right|^2 +
  \nonumber \\ &+&
  {\lambda_{WW} \over 8 \pi} \left|
    \sum_{n=1}^4 {[2(m_\chi/m_W)^2-1] m_{\chi_n}
      \Im(g^2_{Z\chi\chi_n})
      \over m^2_{\chi_n} + m_\chi^2 - m_Z^2 
      -im_{\chi_n}\Gamma_{\chi_n}} 
  \right|^2 ,
\\
  \sigma_{H^+W^-} v &=& \sigma_{H^-W^+} v  = 
  {\lambda_{H^\pm W}^3 m_\chi^2 \over 16 \pi m_W^2} \, \left| 
    \sum_{i=1}^3 { 4 i g_{Wh_iH^\pm} \Im(g_{h_i\chi\chi}) m_{\chi} 
      \over m_{h_i}^2 - 4 m_\chi^2 -im_{h_i}\Gamma_{h_i}} +
\nonumber \right. \\ && + \left.
    \sum_{c=1}^2
    { (g^R_{W\chi \tilde{\chi}^+_c} g^{R*}_{H\chi \tilde{\chi}^+_c} - g^L_{W\chi \tilde{\chi}^+_c} g^{L*}_{H\chi \tilde{\chi}^+_c}) 
      m_{\chi} + 
      (g^L_{W\chi \tilde{\chi}^+_c} g^{R*}_{H\chi \tilde{\chi}^+_c} - g^R_{H\chi \tilde{\chi}^+_c} g^{L*}_{H\chi \tilde{\chi}^+_c}) m_{\tilde{\chi}^+_c}
      \over m^2_{\tilde{\chi}^+_c}+m_\chi^2-(m_{H^\pm}^2+m_W^2)/2
      -im_{\tilde{\chi}^+_c}\Gamma_{\tilde{\chi}^+_c}} 
  \right|^2 ,
\\
  \sigma_{h_iZ} v &=&  
  {\lambda_{h_iZ}^3 m_\chi^2 \over 16 \pi m_Z^2} \, \left| 
    \sum_{n=1}^4
    { - 2 \Re(g_{Z\chi\chi} g^*_{h_i\chi_1\chi_n}) m_{\chi} + 
        2 \Re(g_{Z\chi\chi} g_{h_i\chi_1\chi_n}) m_{\chi_n}
      \over m^2_{\chi_n}+m_\chi^2-(m_{h_i}^2+m_Z^2)/2
      -im_{\chi_n}\Gamma_{\chi_n}}  +
\nonumber \right. \\ && \qquad\qquad + \left.
    \sum_{j=1}^3 { 4 i g_{Zh_jh_i} \Im(g_{h_j\chi\chi}) m_{\chi} 
      \over m_{h_j}^2 - 4 m_\chi^2 -im_{h_j}\Gamma_{h_j}} -
    { g_{h_iZZ} \Re(g_{Z\chi\chi}) \over m_Z^2 }
  \right|^2 ,
\\
  \sigma_{h_ih_j} v &=& {\lambda_{hA} \over 64 \pi m_\chi^2} \left| 
    \sum_{n=1}^4 
    { 4 \Im(g_{h_i\chi_1\chi_n} g^*_{h_j\chi_1\chi_n}) m_\chi m_{\chi_n} +
      (m_{h_i}^2-m_{h_j}^2) \Im(g_{h_i\chi_1\chi_n} g_{h_j\chi_1\chi_n}) 
      \over m^2_{\chi_n} + m_\chi^2 - (m_h^2+m_Z^2)/2 
      -im_{\chi_n}\Gamma_{\chi_n}} + 
    \nonumber \right. \\ && \qquad\qquad + \left.
    \sum_{k=1}^3 { 2 g_{h_ih_jh_k} \Im(g_{h_k\chi\chi}) m_{\chi} 
      \over m_{h_k}^2 - 4 m_\chi^2 -im_{h_k}\Gamma_{h_k}} +
    {i g_{Zh_ih_j} (m_i^2-m_j^2) \Re(g_{Z\chi\chi}) \over m_Z^2 }
  \right|^2 .
\end{eqnarray}
$N_f$ is 3 for quarks and 1 for leptons, $g_{h_iff}$ and $g_{h_i\chi_m\chi_n}$
are given in eqs.~(\ref{h-q-q}--\ref{h-x-x}), $g^L_{\tilde{f}\chi f}$ and
$g^R_{\tilde{f}\chi f}$ are given in eqs.~(\ref{sq-x-q}--\ref{sq-x-q-last}),
and
\begin{eqnarray}
g_{Wh_iH^\pm} &=& {g\over 2} \left[ O'_{i1} \sin(\alpha-\beta) + O'_{i2}
  \cos(\alpha-\beta) + i O'_{i3} \right] , \\
g_{Zh_ih_j} &=& {ig\over2\cos\theta_W} \left[O'_{i1}\sin(\alpha-\beta) +
  O'_{i2} \cos(\alpha-\beta) \right] O'_{j3} - (i \leftrightarrow j) , \\
g^A_{Zff} &=& {g T_{3f} \over 2 \cos\theta_W} , \\
g^L_{W\chi\chi^+_c} &=& - { g N_{14} V^*_{c2} \over \sqrt{2}} + g N_{12}
V^*_{c1} , \\
g^R_{W\chi\chi^+_c} &=& + { g N^*_{13} U_{c2} \over \sqrt{2}} + g N^*_{12}
U_{c1}, \\
g_{Z\chi_m\chi_n} &=& {g \over 2\cos\theta_W} \left( N_{m4} N_{n4}^* - N_{m3}
  N_{n3}^* \right)  .
\end{eqnarray}
Here $V$ and $U$ are the chargino mixing matrices.

For real $\mu$ and real gaugino masses, as we assume here, the terms containing
$\Im(g^L_{W\chi \tilde{\chi}^+_c}g^{R*}_{W\chi \tilde{\chi}^+_c}) $ and
$\Im(g^2_{Z\chi\chi_n})$ in eqs.~(\ref{sww}--\ref{szz}) vanish.

Notice that in the annihilation into fermion pairs, in the first terms under
absolute values in eq.~(\ref{xxff}) (see fig.~1(a)), there can be contributions
from all Higgs bosons $h_i$ for which the imaginary part of $g_{h_i\chi\chi}$
is non-zero. Examining eq.~(\ref{h-x-x}) for the couplings, recalling that the
matrix elements $O'_{ij}$ are real and that for real $\mu$ and real gaugino
masses $N_{1i} N_{1j}$ are also real, we see that the $h_i$ contributes when
$O'_{i3}$ is non-zero. In the CP-conserving case, this happens only for $i=3$,
i.e.\ for the $A$ boson, while with CP violation this occurs also for $i=1$ and
$i=2$. The annihilation into $f\!\bar{f}$ then proceeds through exchange of all
Higgs bosons, raising the possibility of resonant annihilation when $2 m_\chi$
is approximately equal to the mass of any Higgs boson. This phenomenon is
peculiar to CP violation.  An example is given in fig.~9 below.

\section{Results}

\subsection{Results for the elastic scattering cross section}

In Figure 2 we show the neutralino--proton elastic scattering cross section as
a function of neutralino mass for the $\sim 10^6$ values in SUSY parameter
space that we consider.  There is no CP violation in the lower panel ($\Im
A=0$), while CP violation is allowed in the upper panel. Also shown are the
present experimental bounds from the DAMA \cite{dama} and CDMS \cite{cdms}
collaborations as well as the future reach of the CDMS (Soudan) \cite{cdms},
CRESST \cite{cresst} and GENIUS \cite{genius} experiments.  In the upper panel,
it is the maximally enhanced cross section (as a function of $\arg(A)$) that is
plotted. The red (dark) points refer to those values of parameter space which
have the maximum value of the cross section for nonzero $\Im(A)$ and which are
experimentally excluded at zero $\Im(A)$. The blue (grey) region refers to
those values of parameter space which are enhanced when CP violation is
included and which are allowed also at zero $\Im(A)$.  The green (light grey)
empty squares refer to those values of parameter space which have no
enhancement when CP violation is included.  From the existence of the red
points we conclude that there are indeed points in SUSY parameter space which
are ruled out experimentally when CP is conserved but are allowed when CP is
violated.

By comparing corresponding points in the upper and lower panels of figure 2, we
notice that there can be enhancement or suppression of the cross section when
we allow for CP violation. There are two types of enhancement: one in which the
model without CP violation is allowed and another in which it is experimentally
ruled out.  In the first case, it is possible to define a ratio between
enhanced and unenhanced cross sections, $R_{\rm max} = \sigma^{\rm
  max}/\sigma^0_{\rm max}$. in the second case, when both $\sigma(0)$ and
$\sigma(\pi)$ are excluded, it is not possible to define the previous ratio.
Here $\sigma^{\rm max}$ is the maximally enhanced cross section as one goes
through the phase of $A$, and $\sigma^0_{\rm max} =
\max[\sigma(0),\sigma(\pi)]$ is the larger of the unenhanced CP conserving
cross sections.  We plot $R^{\rm max}$ in figure 3 as a function of
$\sigma^0_{\rm max}$.  In those models in parameter space that we have
considered, we notice that the enhancement due to CP violation is at most a
factor of two. 

In figure 3, we have also plotted the ratio $R_{\rm min} = \sigma^{\rm min} /
\sigma^0_{\rm min}$, which is a measure of the maximal suppression of the
neutralino--proton cross section when CP violation is included. Here
$\sigma^{\rm min}$ is the maximally suppressed cross section as one goes
through the phase of $A$, and $\sigma^0_{\rm min} =
\min[\sigma(0),\sigma(\pi)]$ is the smaller of the unenhanced CP conserving
cross sections.  We see that significant suppression of the scattering cross
section, as low as $10^{-7}$, is possible.

In figure 4, we show the dependence of these enhancement and suppression
factors $R_{\rm max}$ and $R_{\rm min}$ on the phase $\phi_A$ of $A$.
The points are plotted at those values of $\phi_A$ at which the maximum
or minimum of the scattering cross section occurs.

\subsection{Results for the annihilation cross section}

We also have obtained values for the neutralino annihilation cross section
$\sigma v$ for the case of CP violation through the phase of $A$.  In figure 5
we show the maximum value of $\sigma v$ obtained as we vary $\phi_A$ as a
function of neutralino mass.  As in the analogous Figure 2 for the scattering
cross section, the upper panel includes CP violation while the lower one does
not.  The distinction between red (dark), blue (grey), and green (light grey)
points is as in Figure 2.

Figure 6 shows the enhancement of the annihilation cross section via the ratio
$R^{\rm ann}_{\rm max} = (\sigma v)^{\rm max}/(\sigma v)^0_{\rm max}$.  We see
that the annihilation cross section can be significantly enhanced for CP
violation with ${\rm Im}(A) \neq 0$, by as much as a factor of $10^6$. In all
models for which we find an enhancement in the annihilation cross section of at
least $10^3$, the enhancement is due to an s-channel resonance with the
exchange of one of the Higgs bosons $h_1$, $h_2$ or $h_3$. See fig.~9 for an
example.

A similar ratio can be constructed for $R^{\rm ann}_{\rm min}$ to show that the
suppression due to CP violation can be roughly a factor of 50. The dependence
of these enhancement and suppression factors $R^{\rm ann}_{\rm max}$ and
$R^{\rm ann}_{\rm min}$ on $\phi_A$ are plotted in figure 7.

\subsection{Phase dependence of the results}

In the four panels in each of the fig.~8--11 we display the behavior of the
scattering cross section $\sigma_{\chi p}$, the annihilation cross section
$\sigma v$, the branching ratio $\mathop{\rm BR}(b\to s\gamma)$, and the
lightest Higgs boson mass $m_{h_1}$ as a function of the phase $\phi_A$ of
$A$. In the third and fourth panels we hatch the regions currently ruled out by
accelerator experiments. In all four panels we denote the part of the curves
that is experimentally allowed by thickened solid lines,
and the part that is experimentally ruled out (as seen e.g.\ in the third
and fourth panels) by thinner solid lines.

For the models shown in figs.~8--11, we give the values of the input parameters
and of the neutralino mass and composition (gaugino fraction
$|N_{11}|^2+|N_{12}|^2$) in table 2.

\begin{table*}[ht]
\label{tab:2}
\begin{center}
\begin{tabular}{| c | r r r r |}
 & JEsp4\_001509 & JE27\_\_004174 & JE28\_\_002656 & JEsp4\_002809 \\ 
 & Fig.~8 & Fig.~9 & Fig.~10  & Fig.~11 \\ \hline 
$\mu$ [GeV] & -331.433 & -271.973 & -234.128 & 958.213 \\
$M_2$ [GeV] & 390.064 & 106.141 & 338.688  & -153.256 \\
$m_A$ [GeV] & 84.2527 & 168.935 & 325.691 & 106.804 \\
$\tan\beta$ & 31.6126 & 4.37629 & 1.80096 & 48.4750 \\
$\widetilde{M}$ [GeV] & 1085.05 & 494.379 & 1856.43 & 890.647 \\
$A/\widetilde{M}$ & 2.71920 & 0.661158 & 1.88819 & -2.05440 \\
\hline
$m_\chi$ [GeV]& 191.46 & 54.95 & 172.4 & 77.04 \\
$|N_{11}|^2+|N_{12}|^2$ & 0.9459 & 0.9806 & 0.9571 & 0.99786
\end{tabular}
\end{center}
\caption{Model 
parameters and neutralino mass and composition (gaugino fraction)
for the examples in figs.~8--11.}
\end{table*}

In the case plotted in fig.~8, the possible phases are bound by the limit on
the $b\to s\gamma$ branching ratio. In the allowed regions, the scattering
cross section at CP-violating phases is suppressed, while the annihilation
cross section is enhanced. The latter takes its maximum allowed value when
the $b\to s\gamma$ limit is reached.

Figure 9 presents another case in which the phase of $A$ is bounded by the
$b\to s\gamma$ branching ratio. Here the scattering cross section is enhanced
by only 2\%, while the annihilation cross section is enhanced by a factor of
$\simeq 222$ at $\phi_A \simeq 0.129\pi$. This is due to a resonant
annihilation of the neutralinos through s-channel exchange of the $h_1$ Higgs
boson (fig.~1(a)), which occurs when $2 m_\chi = m_{h_1}$ (see the lowest
panel). Notice that in the CP conserving case, the s-channel exchange of the
CP-even $h_1$ boson vanishes at $v=0$ because for real $\chi\chi h_1$ couplings
the amplitude is proportional to $\overline{\chi} \chi$ which is zero at $v=0$.
In presence of CP violation, the $\chi\chi h_1$ couplings are in general
complex, and the amplitude contains a contribution from $\overline{\chi}
\gamma_5 \chi$ which does not vanish at $v=0$. So the $h_1$ resonant
annihilation seen in fig.~9 is only possible when CP is violated.

Figure 10 shows a case which is experimentally allowed for all values of the
phase $\phi_A$. The maximum of the scattering cross section takes place at the
CP-conserving value $\phi_A=\pi$ and the minimum at $\phi_A=0$. 
The annihilation cross section on the other hand is enhanced by CP violation,
as can be seen in the second panel. Notice that its maximum occurs at
$\phi_A=0.41\pi$, which is not the point of maximal CP
violation $\phi_A=\pi/2$.

Finally fig.~11 displays an example in which both CP conserving cases are
experimentally excluded while some CP violating cases are allowed.  This is one
of the red (dark) points in fig.~5.  The $\phi_A=0$ case is ruled out by the
bounds on both $\mathop{\rm BR}(b\to s\gamma)$ and the Higgs mass, the
$\phi_A=\pi$ case by only the bound on the Higgs mass.  Notice that the
scattering cross section is of the order of $10^{-6}$ pb, in the region probed
by the direct detection experiments. The annihilation cross section peaks at
$\phi_A=3\pi/4$; notice that again this value is not the point of maximal CP
violation $\phi_A=\pi/2$.

\section{Conclusions}

We have examined the effect of CP violation on the neutralino annihilation and
scattering cross sections, which are of importance in calculations of the
neutralino relic density and of the predicted rates for direct and indirect
searches of neutralino dark matter. Specifically we have considered the case in
which the only CP violating phase in addition to the standard model CKM phase
is in the complex soft trilinear scalar couplings $A$ of the third generation.
This phase affects the squark masses and through radiative corrections
generates a mixing between CP-even and CP-odd Higgs bosons.  This mixing
modifies the neutralino annihilation and scattering cross sections in the
kinematic regimes relevant for dark matter detection. Exploring $\sim 10^6$
points in supersymmetric parameter space with a non-zero phase of $A$, we have
found that: (1) the scattering cross section is generally suppressed, even by 7
orders of magnitude in special cases; (2) the annihilation cross section can be
enhanced by factors of $10^6$ as resonant neutralino annihilation through a
Higgs boson becomes possible at CP-violating values of the phase of $A$. We
have also found cases which are experimentally excluded when CP conservation is
imposed but are allowed when CP conservation is violated. Some of these cases
have neutralino masses and cross sections in the region probed by current dark
matter searches.

\section*{Acknowledgments} 
We would like to thank the Department of Energy for support through the Physics
Department at the University of Michigan, and the Max Planck Institut f\"ur
Physik for support during the course of this work. We thank Apostolos
Pilaftsis, Tobias Hurth, and Antonio Grassi for helpful discussions.
Subsequent to the completion of this work we became aware of partially
overlapping work by Falk, Ferstl, and Olive (hep-ph/9908311).

\newpage
\begin{figure}
\epsfig{file=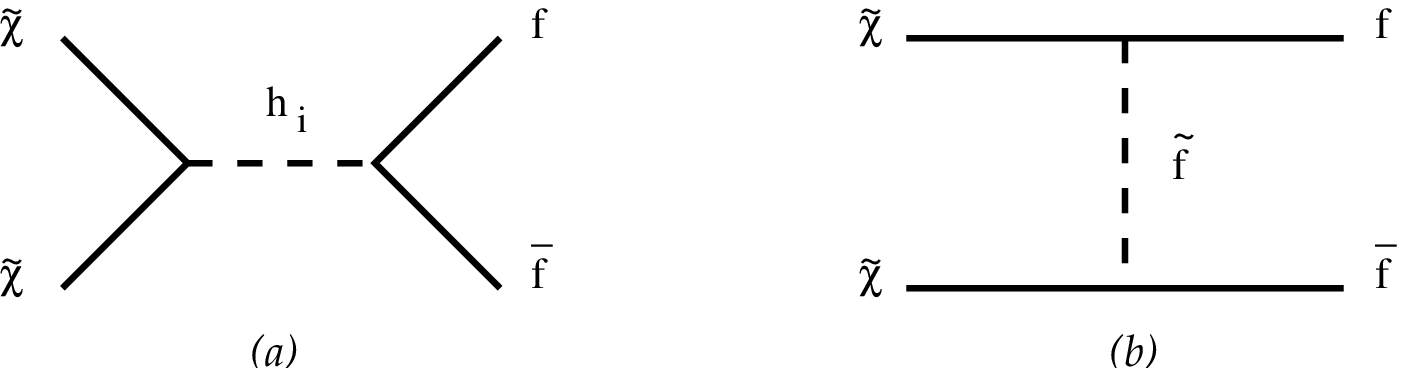,width=\textwidth}
\caption{Processes that contribute to neutralino
  annihilation and scattering and are affected by the CP violating phase of
  $A$. For annihilation: (a) s-channel diagrams via the three neutral Higgs
  bosons $h_1, h_2,$ and $h_3$ into final state fermions $f$, and (b) t-channel
  diagrams via intermediate squarks $\tilde f$ into final state fermions. For
  scattering: crossed diagrams.}
\end{figure}

\newpage
\begin{figure}
\epsfig{file=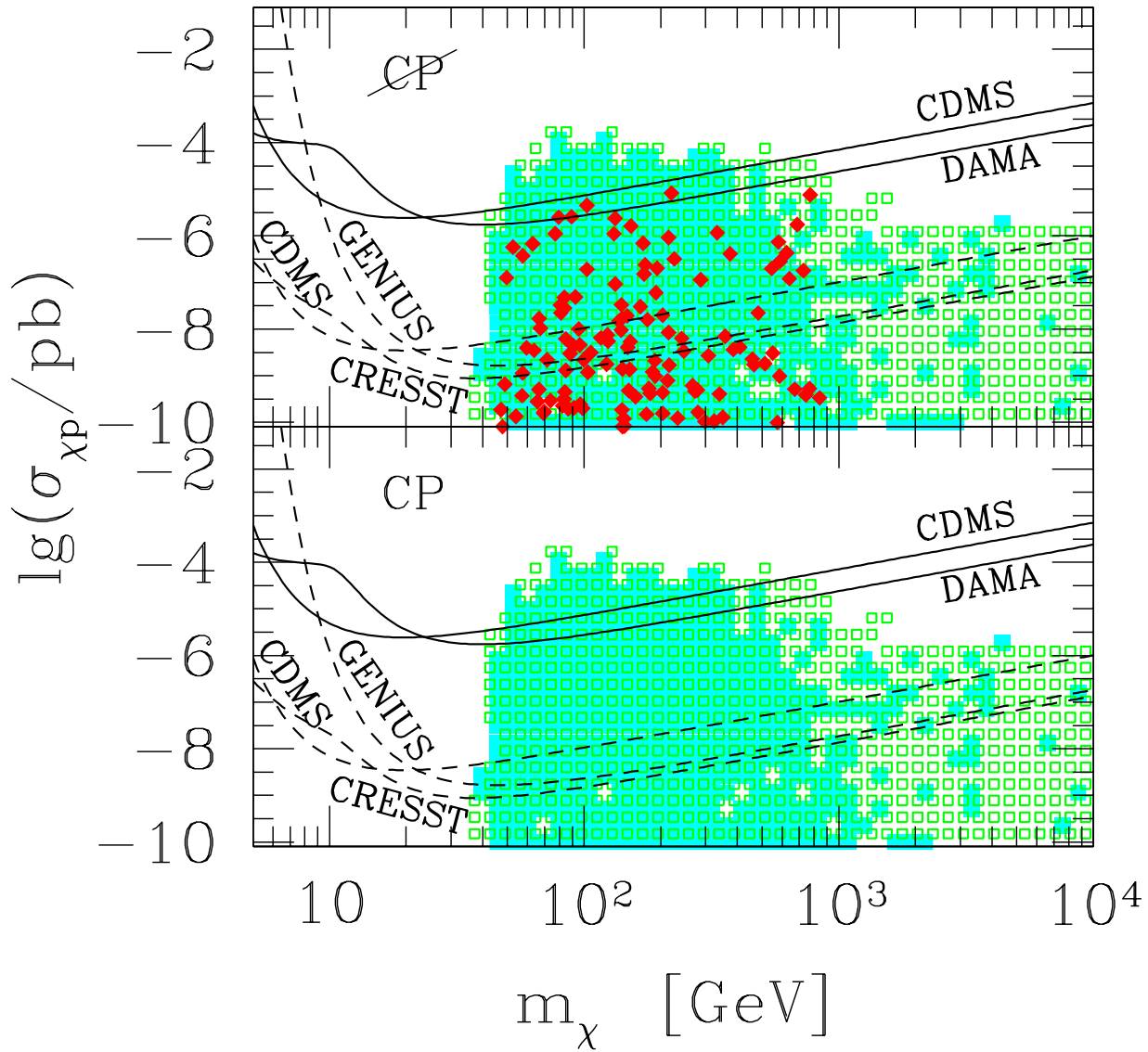,width=\textwidth}
\caption{Neutralino elastic scattering cross section (in
  pb) as a function of neutralino mass (in GeV) for $\sim 10^6$ values in SUSY
  parameter space.  The upper panel is for the case of CP violation via $\Im(A)
  \neq 0$ while the lower panel is for the case of no CP violation.  In the
  upper panel, it is the maximally enhanced cross section (as a function of
  $\arg(A)$) that is plotted. The red (dark) points refer to those values of
  parameter space which have the maximum value of the cross section for nonzero
  $\Im(A)$ and which are experimentally excluded at zero $\Im(A)$. The blue
  (grey) region refer to those values of parameter space which are enhanced
  when CP violation is included and which are allowed also at zero $\Im(A)$.
  The green (light grey) empty squares refer to those values of parameter space
  which have no enhancement when CP violation is included.  The solid lines
  indicate the current experimental bounds placed by DAMA and CDMS; the dashed
  lines indicate the future reach of the CDMS (Soudan), GENIUS, and CRESST
  proposals.}
\end{figure}

\newpage
\begin{figure}
\epsfig{file=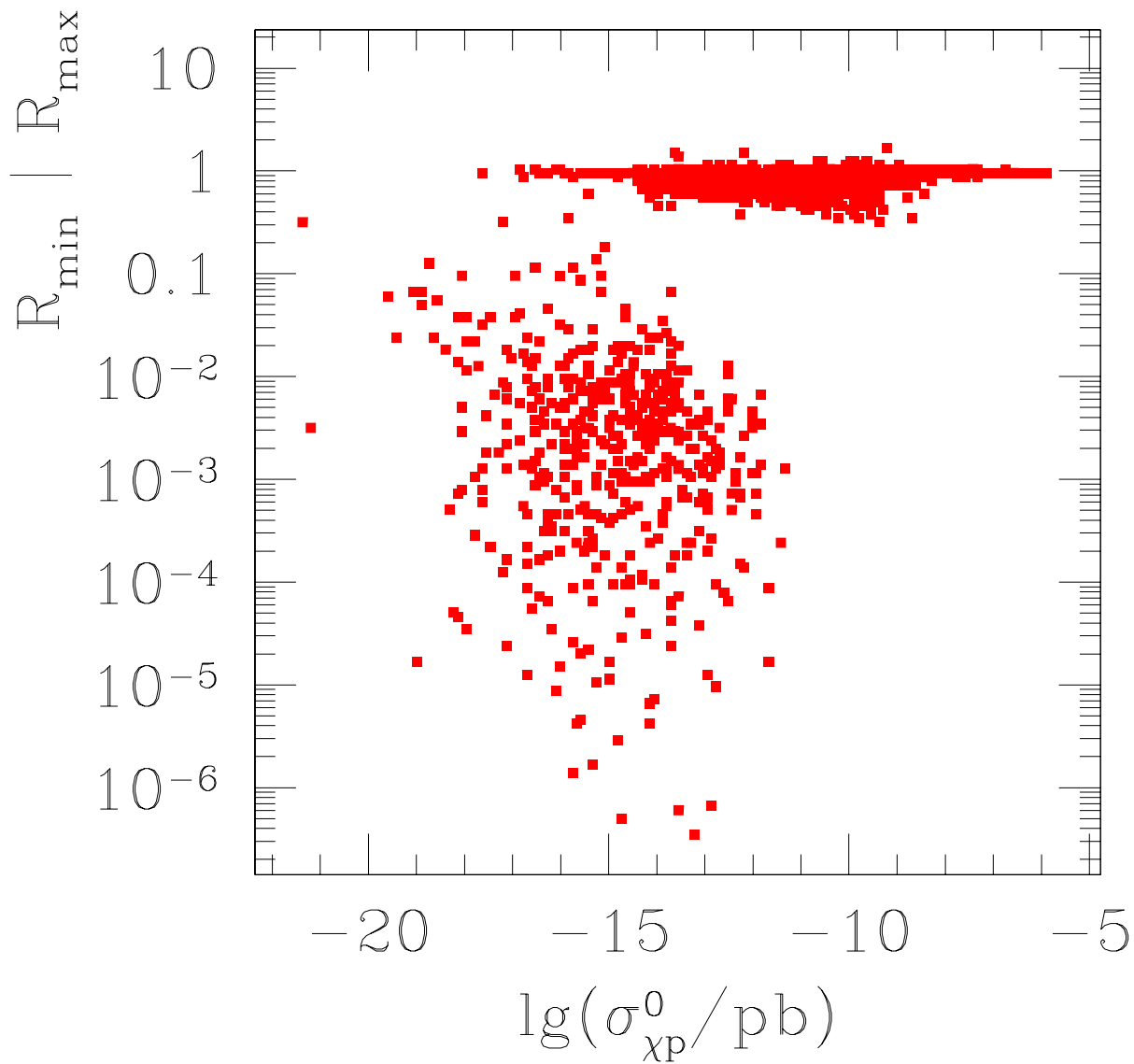,width=\textwidth}
\caption{Enhancement and suppression of elastic
  scattering cross section for the case of CP violating $\arg(A)$.  The plot
  shows the ratio $R_{\rm max} = \sigma^{\rm max}/\sigma^0_{\rm max}$ as a
  function of the unenhanced scattering cross section $\sigma^0_{\rm max}
  =\max[\sigma(0),\sigma(\pi)]$.  Here $\sigma^{\rm max}$ is the enhanced
  scattering cross section and the superscript max indicates the maximal
  enhancement as one goes through the phase of $A$.  The denominator of the
  ratio $R_{\rm max}$ chooses the larger value of the scattering cross section
  without CP violation, i.e., for phase = 0 or phase = $\pi$.  Similarly, the
  ratio $R_{\rm min} = \sigma^{\rm min}/ \sigma^0_{\rm min}$ is plotted; this
  time the denominator chooses the smaller value of the scattering cross
  section without CP violation, $\sigma^0_{\rm min}
  =\min[\sigma(0),\sigma(\pi)]$.}
\end{figure}

\newpage
\begin{figure}
\epsfig{file=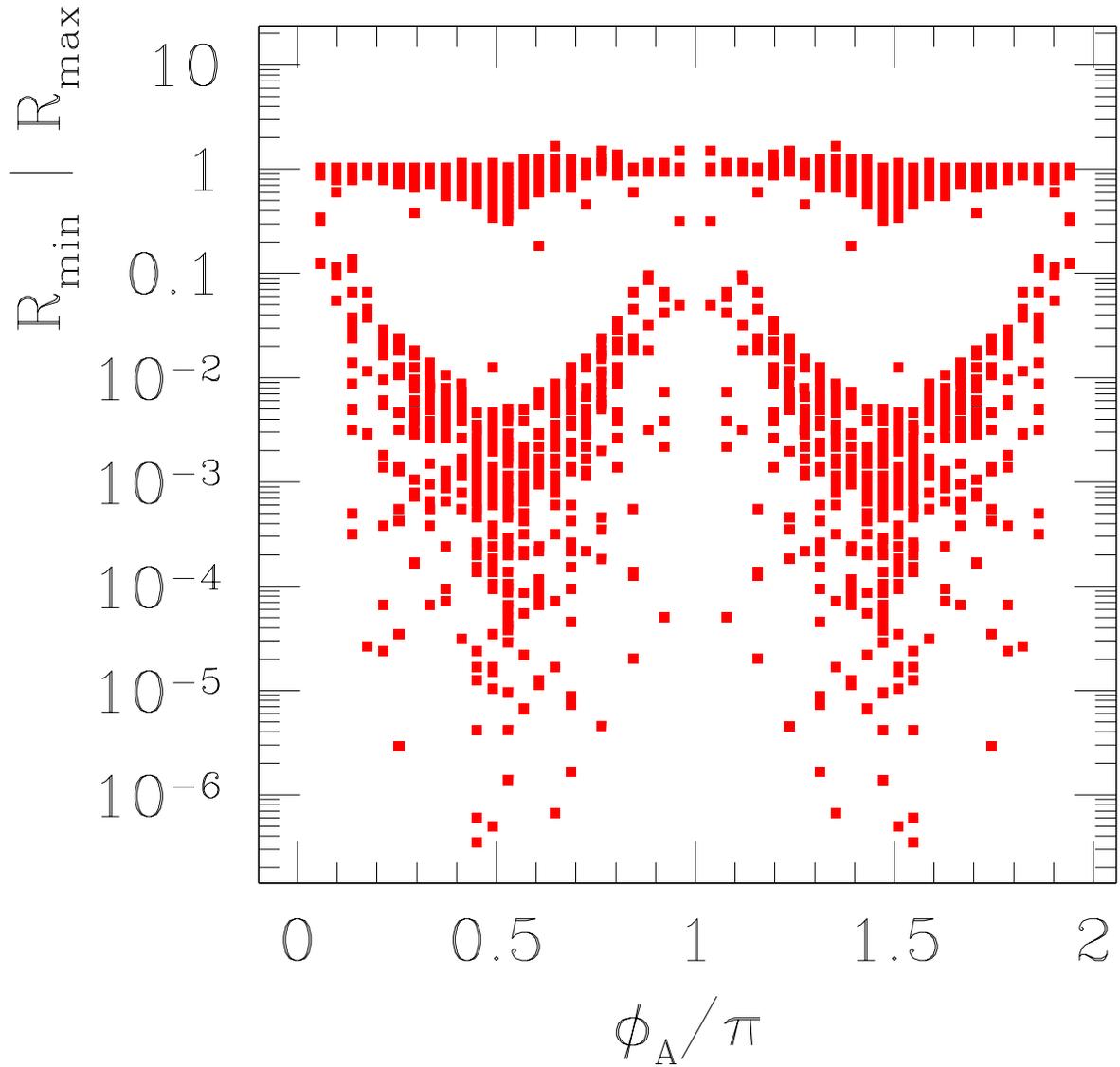,width=\textwidth}
\caption{The enhancement/suppression factors $R_{\rm
    max}$ and $R_{\rm min}$ defined in the caption of figure 3 as
  a function of the values $\phi_A$ of the phase of $A$ where the
  maximum/minimum occur.}
\end{figure}

\newpage
\begin{figure}
\epsfig{file=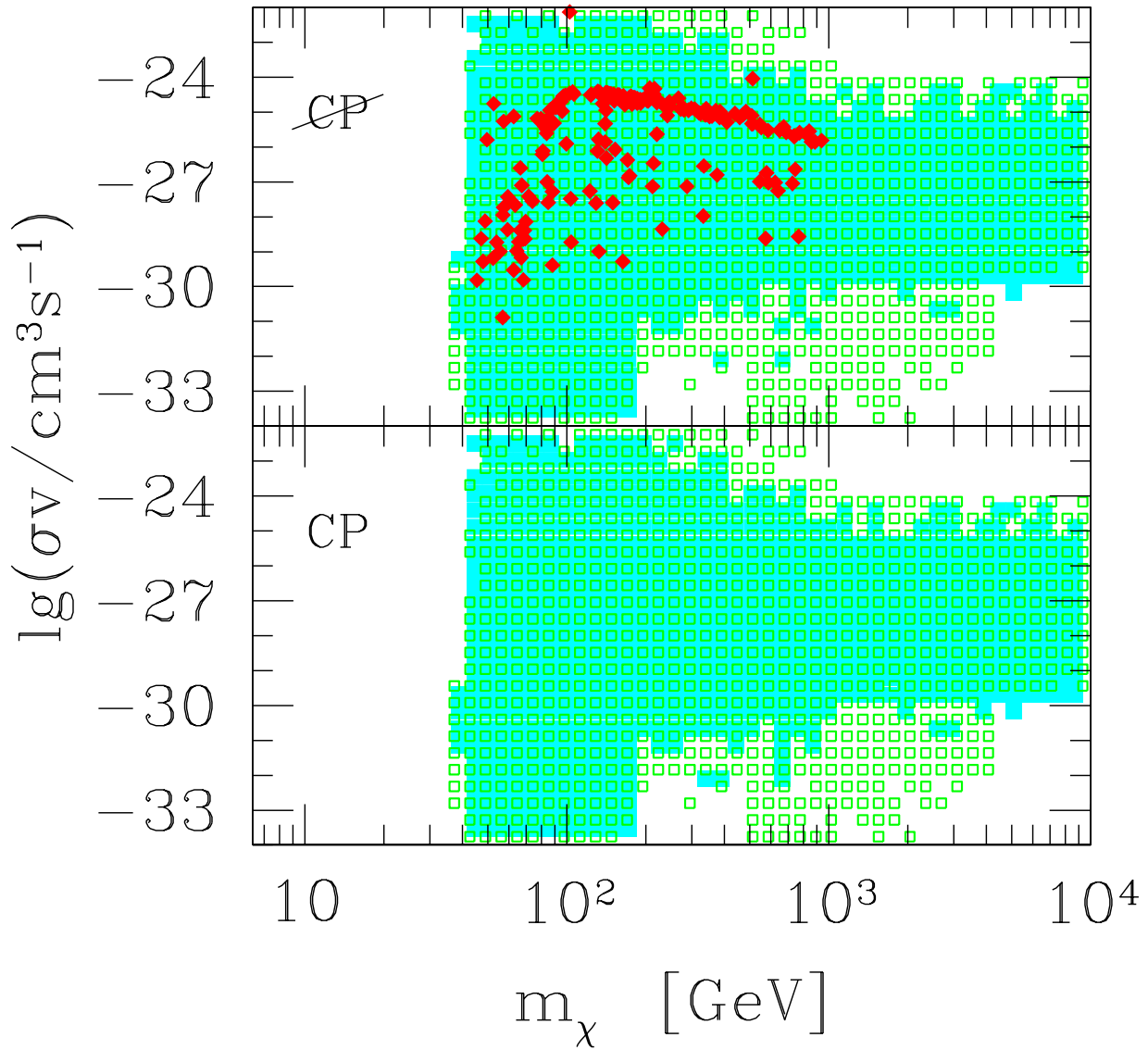,width=\textwidth}
\caption{Same as fig.~2 but for the neutralino annihilation cross
  section times relative velocity $\sigma v$ (in cm$^3$/s at $v=0$) as a
  function of neutralino mass (in GeV) for $\sim 10^6$ values in SUSY parameter
  space. }
\end{figure}

\newpage
\begin{figure}
\epsfig{file=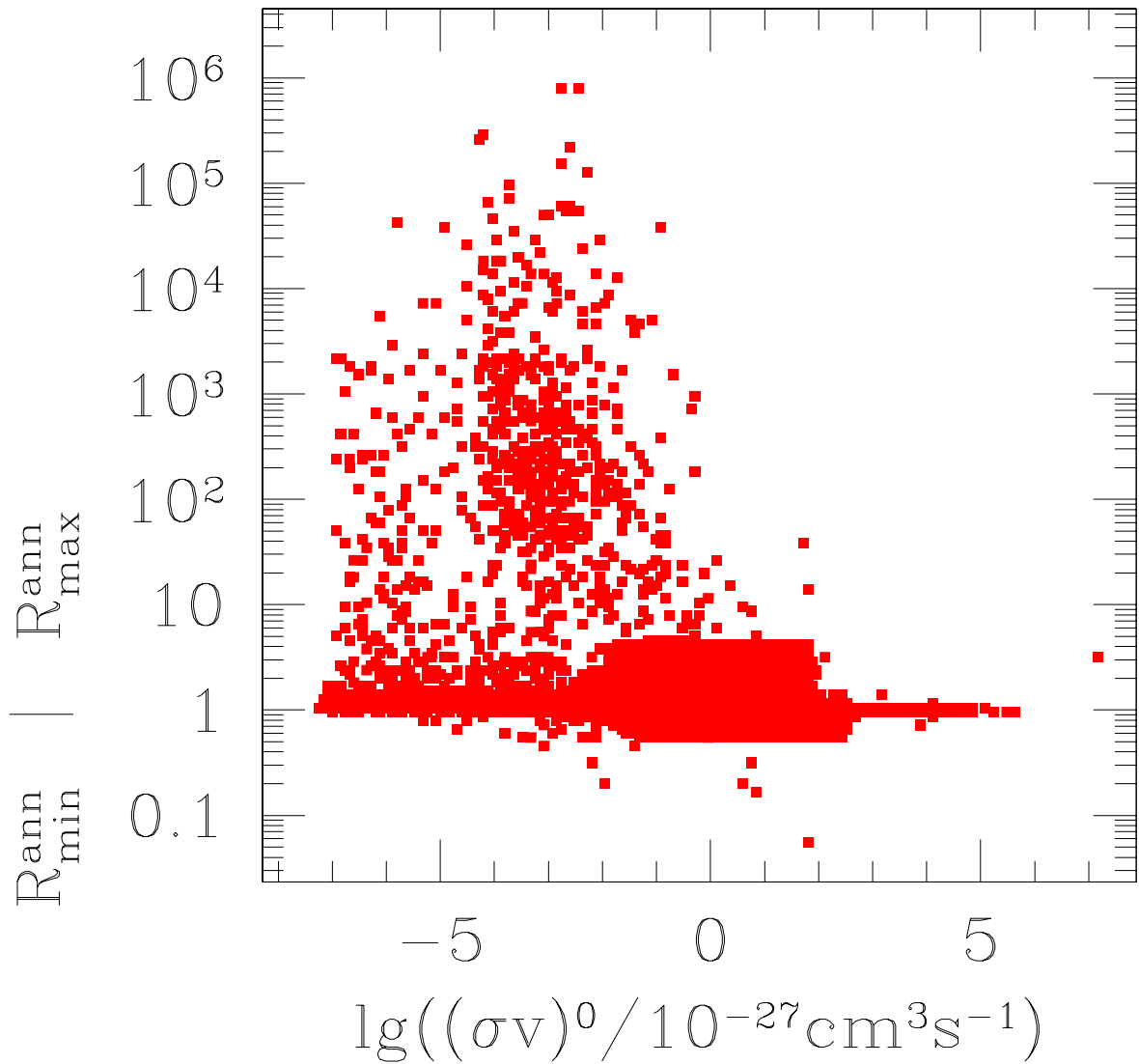,width=\textwidth}
\caption{Enhancement and suppression of neutralino annihilation
  cross section for the case of CP violating $\arg(A)$. The plot shows the
  ratio $R^{\rm ann}_{\rm max} = (\sigma v)^{\rm max}/(\sigma v)^0_{\rm max}$
  as a function of the unenhanced annihilation cross section $(\sigma v)^0_{\rm
    max} =\max[\sigma v(0),\sigma v(\pi)]$.  Here $(\sigma v)^{\rm max}$ is the
  enhanced scattering cross section and the superscript max indicates the
  maximal enhancement as one goes through the phase of $A$.  The denominator of
  the ratio $R^{\rm ann}_{\rm max}$ chooses the larger value of the scattering
  cross section without CP violation, i.e., for phase = 0 or phase = $\pi$.
  Similarly, the ratio $R^{\rm ann}_{\rm min} = (\sigma v)^{\rm min}/(\sigma
  v)^0_{\rm min}$ is plotted.  Here $(\sigma v)^0_{\rm min} =\min[\sigma
  v(0),\sigma v(\pi)]$.  }
\end{figure}

\newpage
\begin{figure}
\epsfig{file=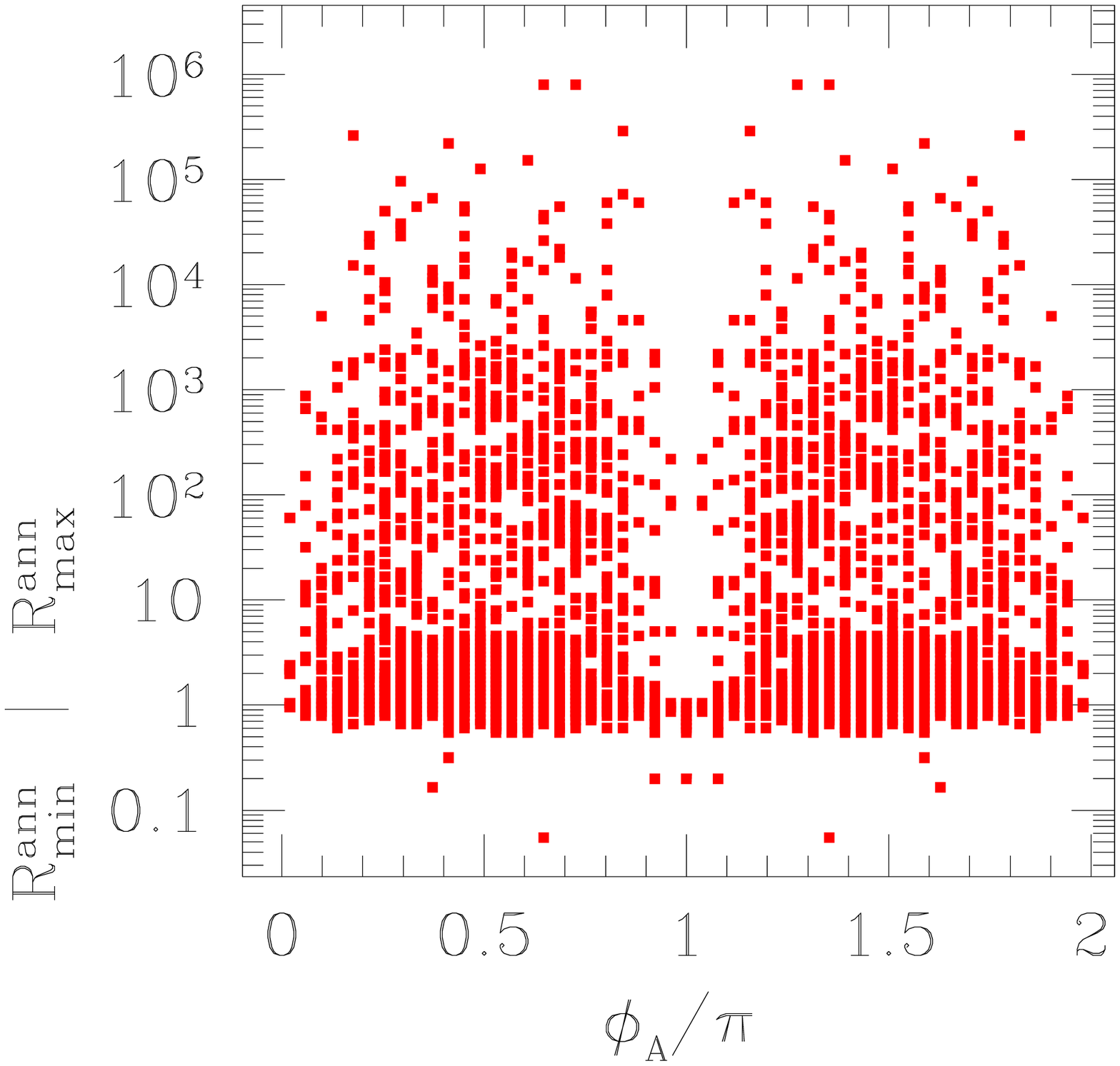,width=\textwidth}
\caption{The enhancement/suppression factors $R^{\rm ann}_{\rm
    max}$ and $R^{\rm ann}_{\rm min}$ defined in the caption of figure
  6 as a function of the values $\phi_A$ of the phase of $A$ where
  the maximum/minimum occur. }
\end{figure}

\newpage
\begin{figure}
\epsfig{file=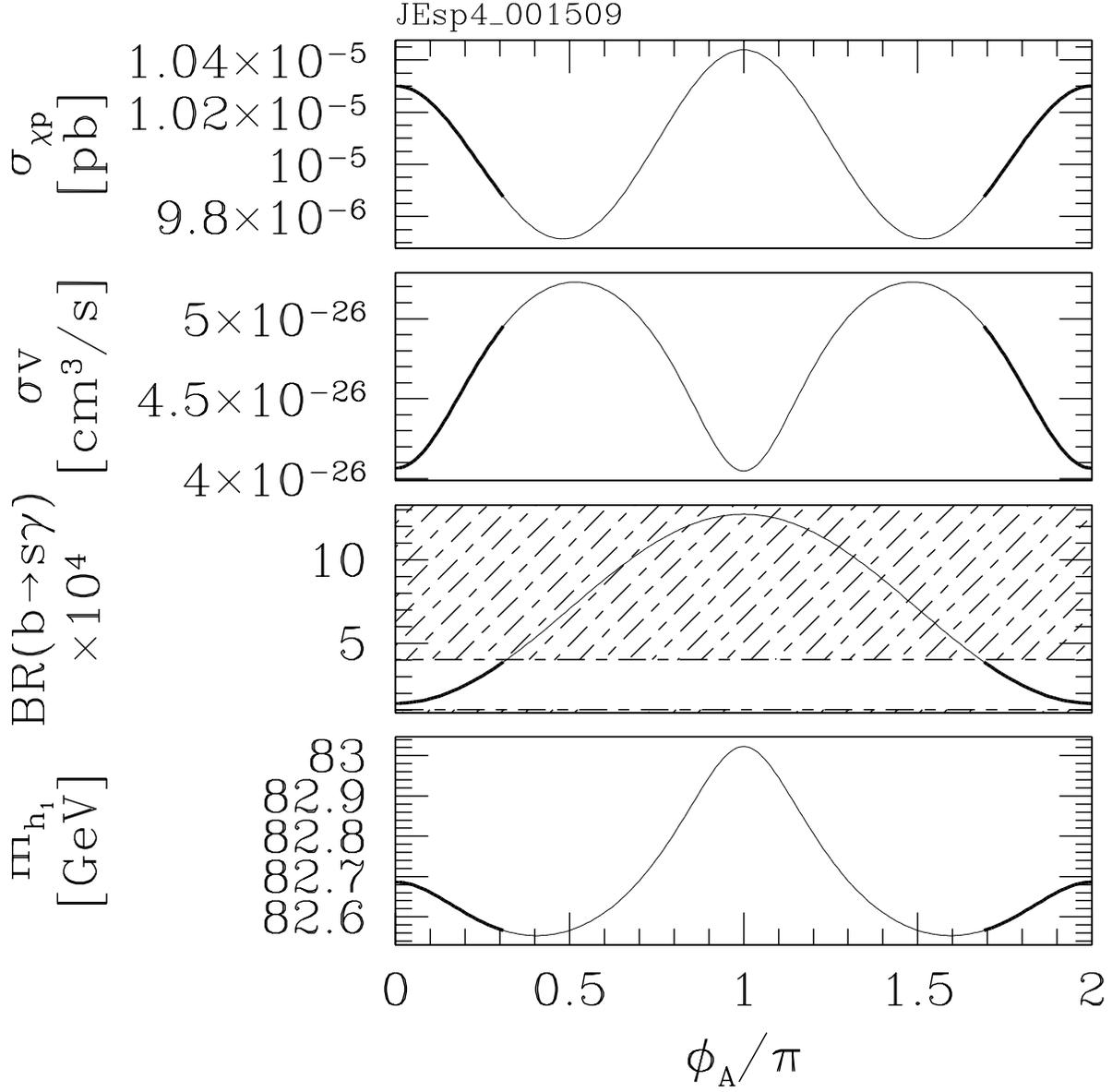,width=\textwidth}
\caption{The four panels from top to bottom display the following:
  the scattering cross section $\sigma_{\chi p}$ in pb, the annihilation cross
  section $\sigma v$ in cm$^3$/s, the branching ratio $\mathop{\rm BR}(b\to
  s\gamma) \times 10^4$, and the lightest Higgs boson mass $m_{h_1}$ in GeV as
  a function of the phase $\phi_A$ of $A$. CP conserving phases are $\phi_A =
  0, \pi$ while all other values are CP violating.  In the third and fourth
  panels we hatch the regions currently ruled out by accelerator experiments.
  In all four panels we denote the part of the curves that is experimentally
  allowed by thickened solid lines, and the part that is experimentally ruled
  out by thinner solid lines.  In this figure, the possible
  phases are bound by the limit on the $b\to s\gamma$ branching ratio. In the
  allowed regions, the scattering cross section at CP-violating phases is
  suppressed, while the annihilation cross section is enhanced. The latter
  takes its maximum allowed value when the $b\to s\gamma$ limit is reached.  }
\end{figure}

\newpage
\begin{figure}
\epsfig{file=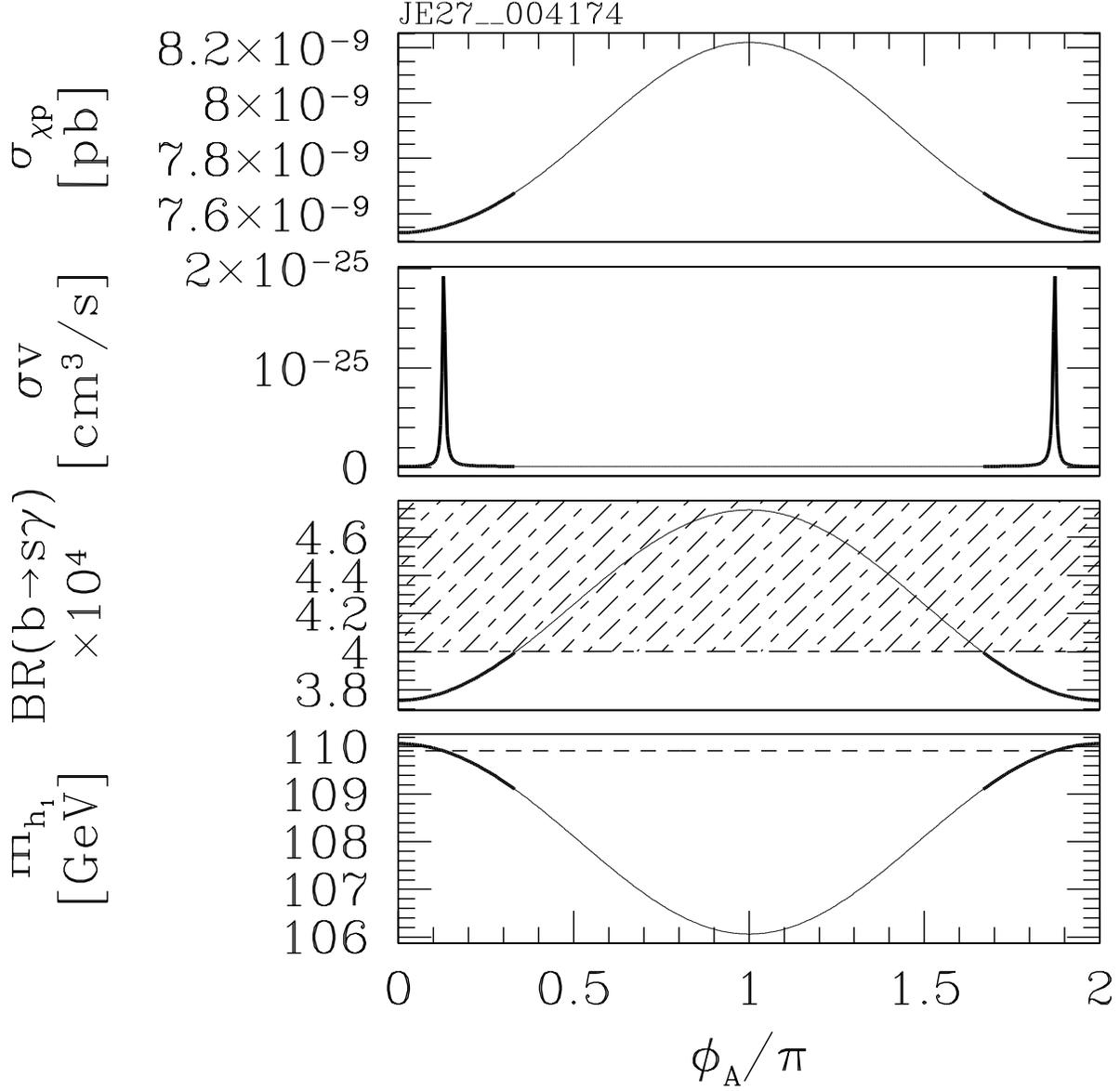,width=\textwidth}
\caption{Same notation as fig.~8.
  The phase of $A$ is bounded by the $b\to s\gamma$ branching ratio, the
  scattering cross section is enhanced by only 2\%, and the annihilation
  cross section is enhanced by a factor of $\simeq 222$ at $\phi_A \simeq
  0.129\pi$, where the annihilation proceeds through the $h_1$ resonance at $2
  m_\chi = m_{h_1}$ (see bottom panel).  }
\end{figure}

\newpage
\begin{figure}
\epsfig{file=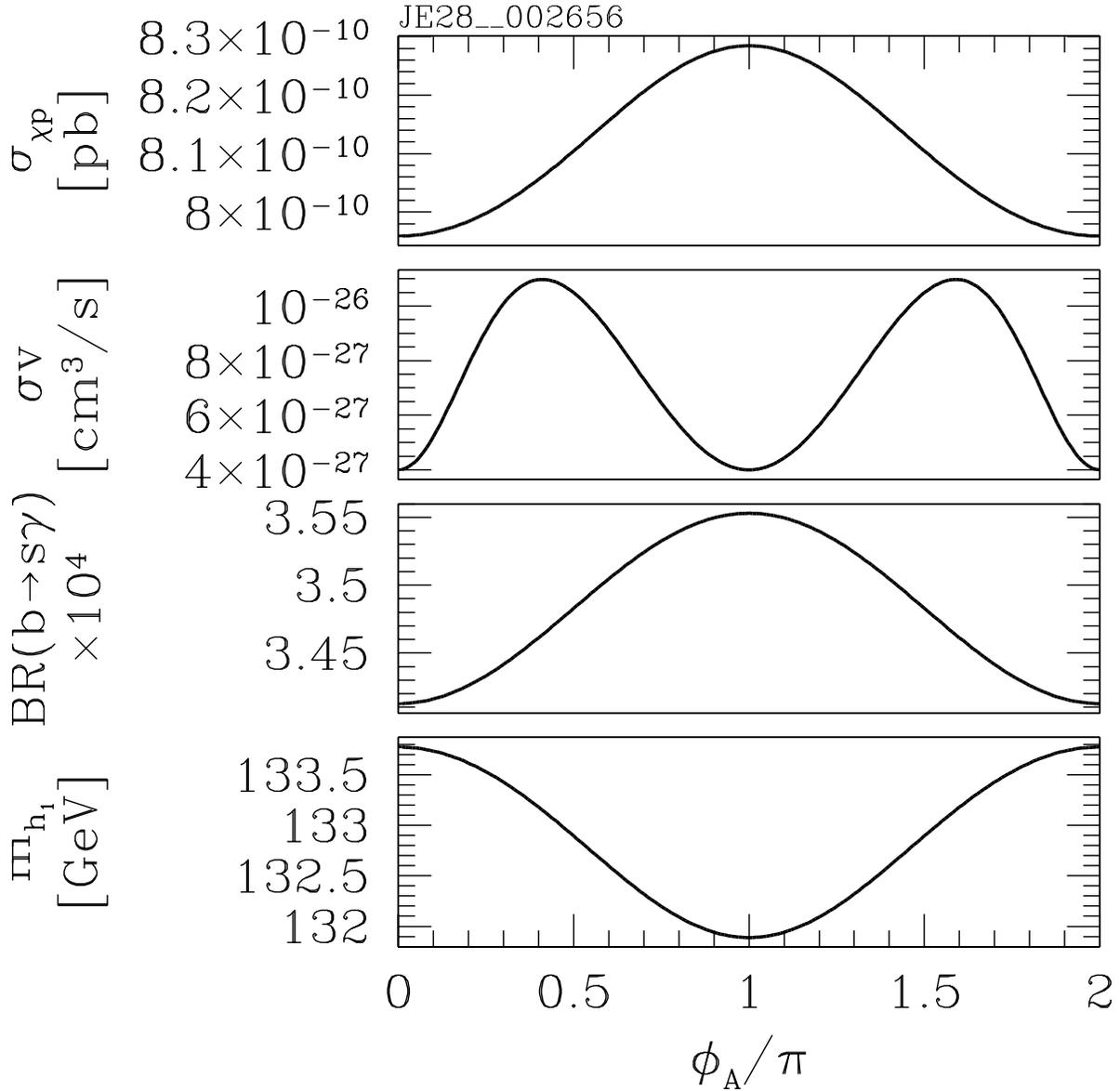,width=\textwidth}
\caption{{\it ``The Duck.''} Same notation as fig.~8. 
  This case is experimentally allowed for all values of the phase $\phi_A$. The
  maximum of the scattering cross section takes place at the CP-conserving
  value $\phi_A=\pi$ and the minimum at $\phi_A=0$. The annihilation cross
  section is enhanced by CP violation, as can be seen in the
  second panel.  }
\end{figure}

\newpage
\begin{figure}
\epsfig{file=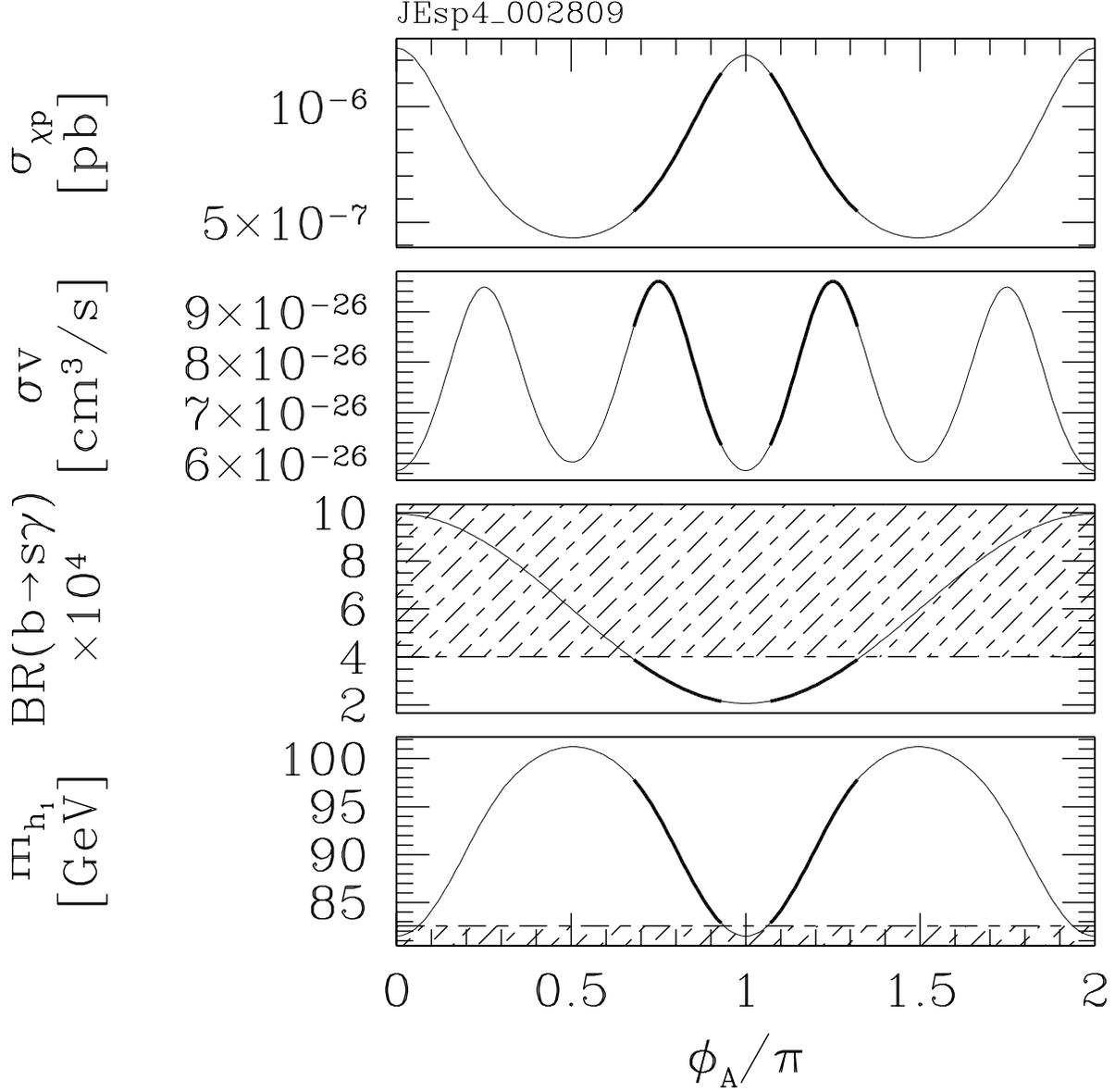,width=\textwidth}
\caption{Same notation as fig.~8.
  Here, both CP conserving cases are experimentally excluded while some CP
  violating cases are allowed.  This is one of the red (dark) points in fig.~5.
  The scattering cross section is of the order of $10^{-6}$ pb, and lies in the
  region being probed by direct detection experiments. The annihilation cross
  section peaks at $\phi_A=3\pi/4$; notice that this value is not the point of
  maximal CP violation $\phi_A=\pi/2$.  }
\end{figure}

\end{document}